\PassOptionsToPackage{numbers,sort&compress}{natbib}
\documentclass[3p]{elsarticle}
\usepackage[latin1]{inputenc}
\usepackage{comment}
\usepackage{amsmath,amssymb,amsfonts}
\usepackage{graphicx}
\usepackage{mathrsfs}
\usepackage{xcolor}
\usepackage{booktabs}  
\usepackage{dcolumn}   
\usepackage{caption}   
\usepackage{booktabs} 
\usepackage{multirow} 
\usepackage{threeparttable} 
\usepackage[normalem]{ulem} 
\usepackage{hyperref}

\let\abs=\envert

\newcommand\opc[1]{{\cal #1}}

\newcommand\ic{{\rm i}}
\newcommand\un[1]{{\rm\,#1}}

\newcommand\numeq[1]{(\ref{#1})}
\newcommand\eq[1]{Eq.~(\ref{#1})}
\newcommand\eqs[1]{Eqs.~(\ref{#1})}
\newcommand\fig[1]{Fig.~\ref{#1}}
\newcommand\numfig[1]{\ref{#1}}
\newcommand\figs[1]{Figs.~\ref{#1}}
\newcommand\sect[1]{Section~\ref{#1}}

\newcommand\sects[1]{Sections~\ref{#1}}
\newcommand\numsect[1]{\ref{#1}}
\newcommand\appen[1]{\ref{#1}}

\newcommand\tabla[1]{Table~\ref{#1}}

\newcommand\cbb[1]{\mathbb{#1}}
\newcommand\wh[1]{\widehat{#1}}

\newcommand\insertfig[6]
{\begin{figure}[#1]
    \begin{center}
        \includegraphics[bb = 0 0 #4cm #5cm]{#3.jpg}
    \end{center}
    \caption{#6}
    \label{#2}
\end{figure}}

\begin{document}

\begin{frontmatter}

\title{An extended method for Statistical Signal Characterization using moments and cumulants, as a fast and accurate pre-processing stage of simple Artificial Neural Networks applied to the recognition of pattern alterations in pulse-like waveforms, suitable for low-resource computational systems}
\author{G.H. Bustos}
\author{H.H. Segnorile\corref{cor1}}
\ead{hector.segnorile@unc.edu.ar}
\cortext[cor1]{Corresponding author}
\address{Facultad de Matem\'{a}tica, Astronom\'{i}a, F\'{i}sica y Computaci\'{o}n (FAMAF),
Universidad Nacional de C\'{o}rdoba,  M. Allende y H. de la Torre - Ciudad Universitaria, X5016LAE - C\'{o}rdoba, Argentina. \\
Instituto de F\'{i}sica Enrique Gaviola - CONICET - C\'{o}rdoba, Argentina.}

\begin{abstract}
We propose a feature-extraction procedure based on the statistical characterization of waveforms, applied as a fast pre-processing stage in a pattern recognition task using simple artificial neural network models. This procedure involves measuring a set of 30 parameters, including moments and cumulants obtained from the waveform, its derivative, and its integral. The technique is presented as an extension of the Statistical Signal Characterization method, which is already established in the literature, and we referred to it as ESSC.\\
As a testing methodology, we employed a procedure to distinguish a pulse-like signal from different versions of itself with altered or deformed frequency spectra, under various signal-to-noise ratio (SNR) conditions of Gaussian white noise. The recognition task was performed by machine learning networks using the proposed ESSC feature extraction method. Additionally, we compared the results with those obtained using raw data inputs in deep learning networks. The algorithms were trained and tested on cases involving Sinc-, Gaussian-, and Chirp-pulse waveforms. We measure accuracy and execution time for the different algorithms solving these pattern-recognition cases, and evaluate the architectural complexity of building such networks.\\
We conclude that a simple multi-layer perceptron network using ESSC can achieve an accuracy of around 90\%, comparable to that of deep learning algorithms, when solving pattern recognition tasks in practical scenarios with SNR above $20\un{dB}$. Additionally, this approach offers an execution time approximately 4 times shorter and significantly lower network construction complexity, enabling its use in low-resource computational systems.
\end{abstract}


\end{frontmatter}

\section{Introduction}\label{sec:intro}

    In pattern recognition, feature extraction is a crucial task in obtaining a higher-level data representation. Various techniques are commonly employed in this process, including the Fast Fourier Transform (FFT), convolution, and Statistics Feature-Based (SFB) methods. These techniques are applied across a wide range of fields, such as computer vision, biomedical signal processing, analog-to-digital communication classification, and fault diagnosis in electrical or mechanical systems, among others. In SFB techniques, manual parameter extraction methods, such as the mean, standard deviation (or variance), $k^{\text{th}}$ moments \cite{Esmael12,Delpha2018} (including kurtosis and skewness \cite{Pakhomov03,Gryllias2012}), or high-order cumulants (HOC) \cite{Zhang20}, are used. These data are commonly analyzed using Support Vector Machine (SVM) algorithms to obtain a classification outcome.

    In early 1992, Hirsch \cite{Book_Hirsch92,Hirsch92} first introduced Statistical Signal Characterization (SSC) as an alternative to computationally intense methods such as the FFT or convolution. The cost of extracting a statistical parameter is proportional to $N_{\!s}$, the number of signal samples. The SSC technique is based on the function's extremum values, yielding only four parameters. It has the disadvantage of performing poorly when there are few extrema. However,  the simplicity and lower operational cost make SSC algorithms attractive for embedded systems or when resources are limited. On the contrary, the FFT provides a complete representation of the signal, and convolution permits matching signals that are shifted or scaled in amplitude; however, the computational cost is of the order of $N_{\!s}\,\log_2(N_{\!s})$ \cite{NumRec_3ed_C12-C13}.
    A. Hossen has well exploited this method in biomedical signals in the ref. \cite{Hossen06}, estimating the SSC parameters of the spectral analysis of the pre-processed and filtered R-R interval in the electrocardiogram (ECG), as a screening technique for short-term identification of patients with obstructive sleep apnea (OSA), setting a threshold value to discriminate with an accuracy of 93\%. In ref. \cite{Hossen14}, the author obtains SSC parameters from electromyography and accelerometer signal spectrum and amplitude variation using the Hilbert transform to discriminate between Parkinson's disease and essential tremor, with efficiencies of 82.5\% and 65.5\%, respectively. In ref. \cite{Hossen07}, SSC parameters are applied as a pre-processing stage, along with Artificial Neural Networks (ANN), to build a robust classification framework for classifying among analog and digital modulation techniques in the presence of noise, achieving efficiencies of up to 83\% and 86\% at an SNR of $3\un{dB}$, respectively.

    In 2012, with the AlexNet \cite{Krizhevsky_2012} architecture and advances in computational hardware, Convolutional Neural Networks (CNNs) became a standard for computer vision classification, marking the creation of a new domain known as Deep Learning (DL). CNN addresses the feature extraction and classification process within a simple algorithmic structure, using raw data as input, thereby avoiding the manual feature extraction and optimization that are often required in traditional Machine Learning (ML) \cite{Yamashita_2018, Kiranyaz_2020}. As an example, in ref. \cite{Kiranyaz_2015}, a one-dimensional (1D) CNN is implemented to classify ECG anomalies, replacing the traditional ML approach, which in this case uses raw heartbeat data as input, achieving superior classification performance with an accuracy of over 98\%; in ref. \cite{Truong2018}, the authors applied CNNs to different intracranial and scalp electroencephalogram (EEG) datasets and proposed a generalized retrospective and patient-specific seizure prediction method, achieving sensitivity greater than 75\% and a false prediction rate below to 0.21/h; in ref. \cite{Cao2025}, the authors introduced a novel approach for emotion recognition employing multi-scale EEG features, denominated as the Dynamic Spatial-Spectral-Temporal Network (DSSTNet), which employs a graph convolutional network (GCN) and outperforms the state-of-the-art methods for emotion recognition; in ref. \cite{Oshea18}, radio signal modulation types are classified utilizing a residual neural network (RNN), a kind of CNN algorithm, outperforming the baseline method based on high-order cumulants and XGBoost classifier \cite{XGBoost_Doc2022}, with rates above 80\% accuracy at $10\un{dB}$ of SNR. Some previous work focuses on extracting features from the signal spectrum, which adds an extra computational load and long-time processes to the signals in the time domain. On the other hand, the DL approach requires a large dataset and specialized hardware due to the high number of parameters to be optimized during training, which in turn leads to complex model architectures.

    In this work, we aim to develop a \emph{fast and systematic feature-extraction method} that facilitates the characterization of pulse-form signals for pattern recognition tasks. Such signal forms are used in NMR (Nuclear Magnetic Resonance) \cite{Bauer84, Warren84, Baum85, Friedrich87, Kessler89, Freeman91, Geen91, Ding98, Freeman98, Siegel2004} or EPR (Electron Paramagnetic Resonance) \cite{Spindler2017} pulse techniques, MRI (Magnetic Resonance Imaging) \cite{Sutherland78, Hutchinson78, Bottomley87} procedures, radar applications \cite{Taylor2000,  LHanning2018, Cho2018}, and communication techniques \cite{Michael2002, Lin2005, Hao2009, Chattopadhyay2009, Mohapatra_Th2009, Mehra2010, Pal2012, Sharma2012, Balsells2012, Gandhi2013, Soto2013, Karar2014, Roy2015, Das2016, Yadav2019}, where shaped-pulse methods are used to perform various actions. In this way, using such a pulse characterization method, a pulse quality measurement can be obtained, which can be used to correct some deviations and achieve an efficient technique.
    Our approach will align with statistical characterization procedures, extending Hirsch's SSC method and employing moments and cumulants derived directly from the raw data. Such a statistical characterization strategy can be used as an efficient pre-processing stage for a simple ANN, using traditional ML algorithms, to complete the recognition procedure rather than a DL approach.
    Additionally, the lower operational cost of statistical algorithms that extract features directly from the pulse-like signal is an advantage for implementing this method in demanding processing control or in low-performance embedded systems.

    The remainder of this paper is organized as follows: \sect{sec:mathapp} reviews key concepts from the literature that inform our mathematical approach. \sect{sec:auxfun} introduces auxiliary functions from the original waveform for parameter extraction. \sect{sec:statpar} discusses characterizing functions using moments and cumulants, while \sect{sec:HSSCmeth} summarizes Hirsch's SSC method. We connect the characterization of moments and cumulants from auxiliary functions with the SSC technique. \sect{sec:app_ESSC_ANN} outlines the practical aspects of the proposed characterization method and its applications. In \sect{sec:def_ESSC}, we present our 30-parameter characterization method, referred to as Extended SSC (ESSC), which includes SSC parameters and cumulants from various auxiliary functions. \sect{sec:preproc_algorithm} describes the necessary signal processing algorithm prior to statistical calculations, culminating in the extraction of ESSC parameters. \sect{sec:data_gen} applies this pre-processing technique as a feature-extraction method for a machine learning network in a pattern recognition task, distinguishing among various pulse-like signal versions. We analyze the performance of different classifiers using ML and DL algorithms in \sect{sec:param_perf}, focusing on accuracy, execution time, and architectural simplicity. After selecting the best-suited method for the defined pulse-like waveform recognition problems, \sect{sec:res_ex-pr} presents the results of the ESSC technique using confusion matrices and evaluates the relevance of ESSC parameters with the ReliefF MATLAB{\circledR} algorithm.
    Finally, \sect{sec:disc_conc} discusses the conclusions drawn from this work.

\section{Mathematical approach}\label{sec:mathapp}

    The statistical characterization procedure used in this work relies on the following premise: \emph{extracting a few representative parameters of a signal by applying statistical calculations to the signal itself or an unambiguous auxiliary signal}.
    In this section, we develop such a premise to put in the same context the characterization of functions using moments or cumulants and Hirsch's SSC method.
    The concepts discussed in this section are well-established in the literature; however, it is worthwhile to revisit and organize them to clarify the general ideas underlying the proposed method of this study.

    When we calculate statistical values from a signal, such as the mean or standard deviation, using its amplitude or time dependence, we perform a weighted sum with a kernel function. Then, such values are associated with the time-domain amplitude distribution of the signal; that is, they characterize the signal form. This section discusses how we can use statistical parameters to provide a comprehensive characterization of the signal and identify the most relevant parameters.

    Different types of statistical parameters can be obtained directly from the signal itself or from an auxiliary function derived from it.
    \emph{The auxiliary functions adequately represent the original signal in a pattern recognition or discrimination problem if they can be unambiguously assigned from the set of signals that we want to discriminate}. With these ideas in mind, we describe in the following sections several ways to obtain auxiliary functions, how these functions are related to a variety of statistical parameters, and how such parameters can be used to expand the function in a power series; also, we will relate Hirsch's SSC method as a procedure to obtain an auxiliary function.

\subsection{Auxiliary functions}\label{sec:auxfun}

    We define an auxiliary signal as a function extracted from the original one using a mathematical method, as shown in \fig{fig:func_aux_param}.
    In that figure, different auxiliary functions $g$ are shown, obtained from the signal $f(t_n)$, where the independent discrete real variable $t_n$ is normalized to range between 0 and 1. Hereafter, we use independent discrete variables; however, the concepts can also be applied to continuous ones.
    For instance, in \fig{fig:func_aux_param}
    \renewcommand{\theenumi}{\alph{enumi}}
    \begin{enumerate}
        \item is shown $g(t_n) = f(t_n)/\sum_n \abs{f(t_n)}$, which is equal to the original function but multiplied by a constant (the inverse of the sum of absolute values), and it can be interpreted as a weight function of the variable $t_n$;
        \item we have $g(t_n) = \abs{f(t_n)}/\sum_n \abs{f(t_n)}$ that is equivalent to a probability density function (PDF) of $t_n$;
        \item the function $g(f)$ is a histogram of the amplitude values of $f$ (within range of $t_n$) normalized as a PDF, where each histogram interval has a width of around 1\% of the whole amplitude range $[-0.22,1]$;
        \item is represented by a series of segments bounded by the extrema of the signal, where the set of amplitude and time intervals $g[m] = (A_m, T_m)$ characterizes the function $f(t_n)$; this procedure is part of Hirsch's SSC method \cite{Book_Hirsch92,Hirsch92}.
    \end{enumerate}

    \insertfig{ht}{fig:func_aux_param}{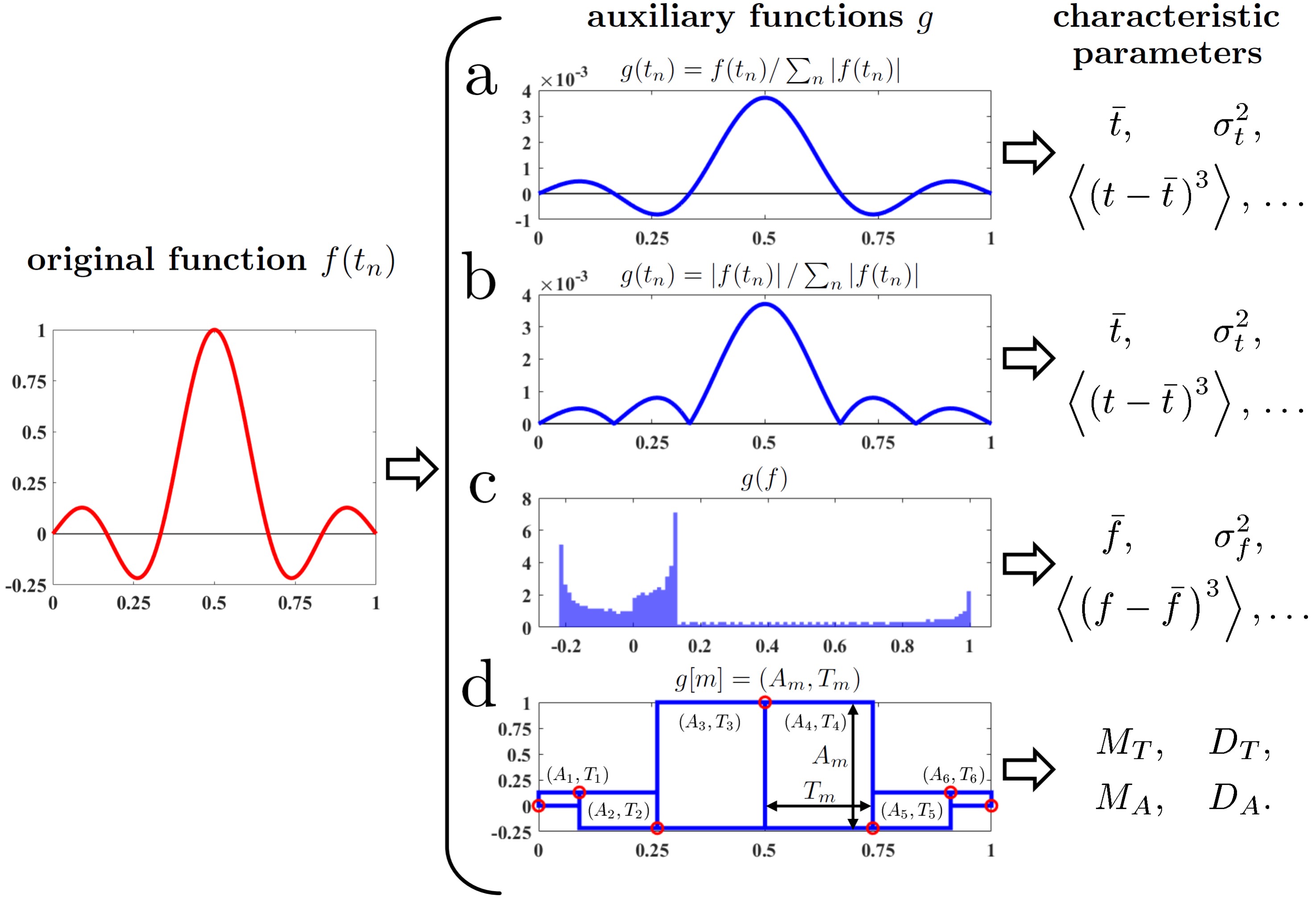}{13.5}{9.3}{Examples of different mathematical methods to obtain auxiliary functions $g$ and characteristic parameters from an original function or signal $f(t_n)$ (the time $t$ is normalized to range between 0 and 1). a: weight function $g(t_n) = f(t_n)/\sum_n \abs{f(t_n)}$; b: probability density function $g(t_n) = \abs{f(t_n)}/\sum_n \abs{f(t_n)}$; c: amplitud histogram $g(f)$ (with an interval width of around 1\% of the whole amplitude range $[-0.22,1]$); d: $g[m] = (A_m, T_m)$ is the set of amplitud and time segments bounded by the extrema of the signal (SSC method developed by Hirsch).}

\subsection{Statistical parameters and signal characterization}\label{sec:statpar}

    To characterize $f(t_n)$, we could extract several parameters from its auxiliary function $g$. Accordingly, we may apply the discrete-time Fourier transform expression \cite{Oppenheim_2ed} to the functions in \fig{fig:func_aux_param} (a) and (b), given by
    \begin{equation}\label{eq:Fourier_dev}
        \wh{g}\,(\omega) \equiv \sum_n g(t_n)\,e^{-\ic\,\omega\,t_n}
        = \sum_{q=0}^{\infty}\frac{\left(-\ic\,\omega\right)^q}{q!}\,\left<t^q\right>,
    \end{equation}
    where $t_n \equiv n\,\Delta t$ is the independent discrete variable, with $n \in \cbb{N}_0$, and $\Delta t$ is the sampling step time of the function $g$.
    In \eq{eq:Fourier_dev}, we defined the $q$-th order moments
    \begin{equation}\label{eq:moment_tq}
    \left<t^q\right> \equiv \sum_n g(t_n)\,t_n^q = \lim_{\omega\rightarrow\,0} \ic^q \frac{d^{\,q}\wh{g}\,(\omega)}{d\omega^q},
    \end{equation}
    with $d^{\,0}\wh{g}\,(\omega)/d\omega^0 \equiv \left<t^0\right> = \sum_n g(t_n) = \wh{g}\,(0)$.
    Then, to obtain $g(t_n)$ from \numeq{eq:Fourier_dev} we can use the inverse transform
    \begin{equation}\label{eq:InvFourier_dev}
        g(t_n) = \frac{\Delta t}{2\pi}\int_{-\pi/\Delta t}^{\pi/\Delta t}d\omega\;\wh{g}\,(\omega)\,e^{\,\ic\,\omega\,t_n}.
    \end{equation}
    On the other hand, \eq{eq:Fourier_dev} can be written as the following expansion \cite{Reichl_2ed_C4}
    \begin{equation}\label{eq:cumulant_exp}
        \wh{g}\,(\omega) = \exp\left[\,\sum_{q=1}^{\infty}\frac{\left(-\ic\,\omega\right)^q}{q!}\;c_q\right] + \left<t^0\right> - 1,
    \end{equation}
    where $c_q$ are the cumulants whose first values are defined as follows
    \begin{equation}\label{eq:cumulant_def}
    \begin{split}
        c_1 &\equiv \left<t\right> \equiv \bar{t},\qquad
        c_2 \equiv \left<t^2\right>-\left<t\right>^2 = \left<\left(t-\bar{t}\,\right)^2\right> \equiv \sigma_t^2,\\
        c_3 &\equiv \left<t^3\right> - 3\left<t\right>\left<t^2\right> + 2\left<t\right>^3 = \left<\left(t-\bar{t}\,\right)^3\right>,\\
        c_4 &\equiv \left<t^4\right> - 4\left<t\right>\left<t^3\right> - 3\left<t^2\right>^2 + 12\left<t\right>^2\left<t^2\right> - 6\left<t\right>^4
            = \left<\left(t-\bar{t}\,\right)^4\right> - 3\left<\left(t-\bar{t}\,\right)^2\right>^2,
    \end{split}
    \end{equation}
    with $\bar{t}$ and $\sigma_t$ as the mean value and the standard deviation of the variable $t_n$ under the weight function or kernel $g(t_n)$.
    In \eq{eq:cumulant_def}, we wrote the cumulants as functions of the moments \numeq{eq:moment_tq} or the $q$-th order central moments
    \begin{equation}\label{eq:moment_tq_cent}
        \left<\left(t-\bar{t}\,\right)^q\right> \equiv \sum_n g(t_n)\,\left(t_n-\bar{t}\,\right)^q.
    \end{equation}

    Therefore, we can use the cumulants \numeq{eq:cumulant_def}, or the central moments \numeq{eq:moment_tq_cent}, or the moments \numeq{eq:moment_tq} to define a set of characteristic parameters representing the original function, as usually found in the bibliography.
    This representation of the signal is important because such parameters are associated with the general properties of the functions.
    For instance, as it is known, the first moment or mean $\bar{t}$ quantifies the average time where the amplitudes are equally distributed around it; the second central moment or variance $\sigma_t^2$ quantifies the spread out of the amplitudes from $\bar{t}$; the third central moment or skewness
    $\left<\left(t-\bar{t}\,\right)^3\right>$ quantifies the lopsidedness of the amplitude distribution values; the fourth central moment or kurtosis
    $\left<\left(t-\bar{t}\,\right)^4\right>$ quantifies the heaviness of the tail (values far from $\bar{t}\,$) of the amplitude distribution values, and so on.
    Moreover, we can approximate the auxiliary function keeping only the terms with lower-order moments or cumulants in \numeq{eq:Fourier_dev} and \numeq{eq:cumulant_exp} for small values of $\omega$ (or for values close to the multiples of $\pm 2\pi/\Delta t$). Hence, the lower frequency part of the spectrum of the auxiliary functions (and for frequencies  $\omega \simeq \pm 2\pi\,m/\Delta t$, $m \in \cbb{N}_0$) is characterized by the lower-order parameters of \eq{eq:cumulant_def}, and we could use such parameters as a tool to identify or discriminate different signals. For example, in the particular case of a gaussian signal, the only non-null cumulants are $c_1 \equiv \bar{t}$ and $c_2 \equiv \sigma_t^2$ (with $\left<t^0\right> = 1$), and we can completely characterize this function with these two parameters.\\

    To finalize this section, we describe other parameters that could be used to characterize the function $f(t_n)$.
    In such a way, we may define the $q$-th order amplitude central moments
    \begin{equation}\label{eq:moment_fq_cent}
        \left<\left(f-\bar{f}\,\right)^q\right> = \frac{1}{N}\sum_n \left(f(t_n)-\bar{f}\,\right)^q,
    \end{equation}
    with the mean value $\bar{f} \equiv \frac{1}{N_{\!s}}\sum_n f(t_n)$ and $N_{\!s}$ is the number of time samples $t_n$. Besides, we have the variance
    $\sigma_f^2 \equiv \left<\left(f-\bar{f}\,\right)^2\right>$.
    Suppose the amplitude range of $f(t_n)$ is divided into segments of dimension $\Delta f$ and we define $\alpha(f)$ as the number of samples that have values between $f$ and $f+\Delta f$. In that case, we could approximate
    \begin{equation}\label{eq:moment_fq_cent_appro}
        \left<\left(f-\bar{f}\,\right)^q\right> \simeq \frac{1}{N_{\!s}}\sum_f \alpha(f)\left(f-\bar{f}\,\right)^q
        = \Delta f\,\sum_f g(f)\left(f-\bar{f}\,\right)^q,
    \end{equation}
    where the sum is performed over the amplitude initial values of the segments.
    In \numeq{eq:moment_fq_cent_appro}, we defined $g(f) \equiv \alpha(f)/(N_{\!s}\,\Delta f)$ as representing a probability density function (PDF) of the amplitude of $f(t_n)$. The smaller the value of $\Delta f$, the better the approximation \numeq{eq:moment_fq_cent_appro} will be.\\
    The expression \numeq{eq:moment_fq_cent_appro} permits the link between the amplitude central moments \numeq{eq:moment_fq_cent} and $g(f)$, similarly to that in \eq{eq:cumulant_exp}. Thus, the Fourier transform of $g(f)$ can be developed in such amplitude moments.
    Then, $g(f)$ could represent an auxiliary function of $f(t_n)$, where the moments \numeq{eq:moment_fq_cent} and the mean value $\bar{f}$ characterize $f(t_n)$, as is shown in \fig{fig:func_aux_param} (c).\\

\subsection{Hirsch's SSC method}\label{sec:HSSCmeth}

    As a final analysis of signal characterization procedures, we present the SSC method developed by Hirsch \cite{Book_Hirsch92,Hirsch92}.
    In this method, we obtain a series of segments bounded by the signal's extrema, as shown in \fig{fig:func_aux_param} (d) (circles represent the extrema). Then, we extract the amplitudes $A_m$ and the time intervals $T_m$ of the $m$-th segment and calculate the following set of mean values or SSC parameters
    \begin{equation}\label{eq:SSC_Hirsch}
    \begin{split}
        &\qquad M_A = \sum_{m=1}^{N_{\!M}} A_m/N_{\!M},\qquad
        M_T = \sum_{m=1}^{N_{\!M}} T_m/N_{\!M},\\
        &D_A = \sum_{m=1}^{N_{\!M}} \abs{A_m - M_A}/N_{\!M},\quad
        D_T = \sum_{m=1}^{N_{\!M}} \abs{T_m - M_T}/N_{\!M},
    \end{split}
    \end{equation}
    where $N_{\!M}$ is the total number of segments of the function.\\
    We can analyze this method similar to the other characterization procedures presented in \fig{fig:func_aux_param}.
    Accordingly, we could consider the set of segments $g[m] = (A_m, T_m)$ as an auxiliary function, with parameters related to the first and central second moments of the amplitude and time segments.
    On the other hand, we could rebuild the original signal by interpolating of the extrema, for example, using linear, polynomial, or sinusoidal functions. Thus, such a set of extrema contains relevant information about $f(t_n)$.
    Moreover, variations in amplitude and time segments are associated with amplitude and frequency modulation, respectively. Consequently, the SSC parameters characterize these modulation behaviors.\\

    Finally, we conclude that a signal can be characterized by the statistical moments (\eqs{eq:moment_tq}, \numeq{eq:moment_tq_cent} or \numeq{eq:moment_fq_cent}), the cumulants \numeq{eq:cumulant_def}, or the SSC parameters \numeq{eq:SSC_Hirsch}.
    The precision of such signal characterization depends on the number and relevance of these parameters (for instance, the influence of a particular moment in \numeq{eq:Fourier_dev} or \numeq{eq:cumulant_exp}).
    The calculation of these statistical parameters involves a sum or an operation over the $N_{\!s}$-sampled values of $f(t_n)$); to compute the SSC parameters, it is necessary to examine all samples to detect the extrema of the function.
    That means their computational strength is $\propto N_{\!s}$, which is very light compared to other techniques, such as convolution, the Discrete Fourier Transform (DFT), or the Fast Fourier Transform, which require a strength $\propto N_{\!s}^2$ or $\propto N_{\!s} \log_2(N_{\!s})$ \cite{Book_Hirsch92}.
    Therefore, these statistical methods are suitable for applications in embedded systems with limited computational capability or in real-time systems, where a fast procedure is required to characterize different signals.\\

    In the next section, we apply the statistical characterization method to a pattern recognition problem, utilizing it as a pre-processing stage in Artificial Neural Networks (ANNs).\\

\section{Application to pattern recognition using neural networks}\label{sec:app_ESSC_ANN}

    In this section, we present a specific set of statistical parameters extracted from a signal, which serves as a distinct method for characterizing and distinguishing the original form from various altered or deformed versions. This approach can be applied to a wide range of signals. The set of parameters we propose will serve as a \emph{fingerprint} of the signal, allowing us to quantify the extent to which its form deviates from the ideal. Accordingly, we can apply the statistical characterization methods discussed in \sect{sec:mathapp} as a fast pre-processing stage for an ANN used in a pattern recognition task. In particular, we aim to characterize and discriminate different pulse-like signals, which are waveforms of bounded time duration used in various electronic applications, such as radar systems, NMR techniques, MRI, and pulse shaping in telecommunications. Moreover, we want to distinguish between a pulse signal and its various versions, which may exhibit amplitude alterations or frequency-spectrum deformations. These pulse deformations are due to imperfections in the amplifiers or the communication channel; therefore, we can adjust the original signal to compensate for these distortions.
    Accordingly, in the next section, we define the parameter set to characterize the different classes of waveforms.

\subsection{Definition of the ESSC parameters}\label{sec:def_ESSC}

    Exhaustive training of an ANN is crucial for achieving success in applying it to solve a task. Thus, a fast and proper pre-processing strategy is necessary, particularly for real-time processing and when the network is used in embedded systems with memory restrictions or limitations.
    Therefore, we use the method of \sect{sec:mathapp} for the auxiliary functions in \fig{fig:func_aux_param} (a) and (c), keeping only the first three cumulants \numeq{eq:cumulant_def}; i.e., we characterize a pulse waveform with its mean value, variance, and skewness of the time and amplitude distribution of $f(t_n)$. Additionally, we incorporate the SSC parameters of \fig{fig:func_aux_param} (d) to characterize the amplitude and frequency modulation of the waveform. Accordingly, using a similar nomenclature to that in \eq{eq:SSC_Hirsch}, we define the set of ten statistical parameters:
    \begin{equation}\label{eq:stat_param}
    \bigg\{M_T,\; D_T,\; M_A,\; D_A,\;
    M_f^1 \equiv \bar{f},\; D_f^2 \equiv \sigma_f^2,\; D_f^3 \equiv \left<\left(f-\bar{f}\,\right)^3\right>,\;
    M_t^1 \equiv \bar{t},\; D_t^2 \equiv \sigma_t^2,\; D_t^3 \equiv \left<\left(t-\bar{t}\,\right)^3\right>\bigg\}.
    \end{equation}
    We could also extract the high- and low-frequency behavior of $f(t_n)$ by means of its derivative or integral, respectively.
    These operation are simply defined as the difference and the accumulation:
    \begin{equation}\label{eq:dis_der_int}
    \begin{split}
    f_D(t_n) &\equiv f(t_{n+1}) - f(t_n), \;\;\text{with}\;\; f_D(t_{N_{\!s}+1}) = 0,\\
    f_I(t_n) &\equiv f_I(t_{n-1}) + f(t_n), \;\;\text{with}\;\; f_I(t_0) = 0,
    \end{split}
    \end{equation}
    for discrete variables ($n \in [1,N_{\!s}]$). Then, the strength to compute the derivative $f_D$ and the integral $f_I$ is $\propto N_{\!s}$; thus, we could incorporate them as auxiliary functions to extract parameters and characterize $f$ without an excessive computational effort.\\
    In this way, we propose to use the set of parameters \numeq{eq:stat_param} calculated from the functions $f$, $f_D$, and $f_I$, as an statistical characterization of $f(t_n)$. Therefore, that 30-parameter set can be considered a \emph{fingerprint} of the signal or the original function, and we name this method Extended Statistical Signal Characterization (ESSC), which is proposed as an extension of Hirsch's SSC. The pre-processing stage of the ANN, using the ESSC method for our pattern recognition problem, is illustrated in \fig{fig:pre-proc}b and explained in \sect{sec:preproc_algorithm}.\\

    In the following sections, we evaluate the performance of ESSC parameters in various machine learning algorithms and compare the results with those of deep learning models to select and train the most suitable ANN to distinguish between an original pulse and other pulses with altered frequency spectra.

\subsection{Pre-processing algorithm with Extended Statistical Signal Characterization}\label{sec:preproc_algorithm}

    \insertfig{h!}{fig:pre-proc}{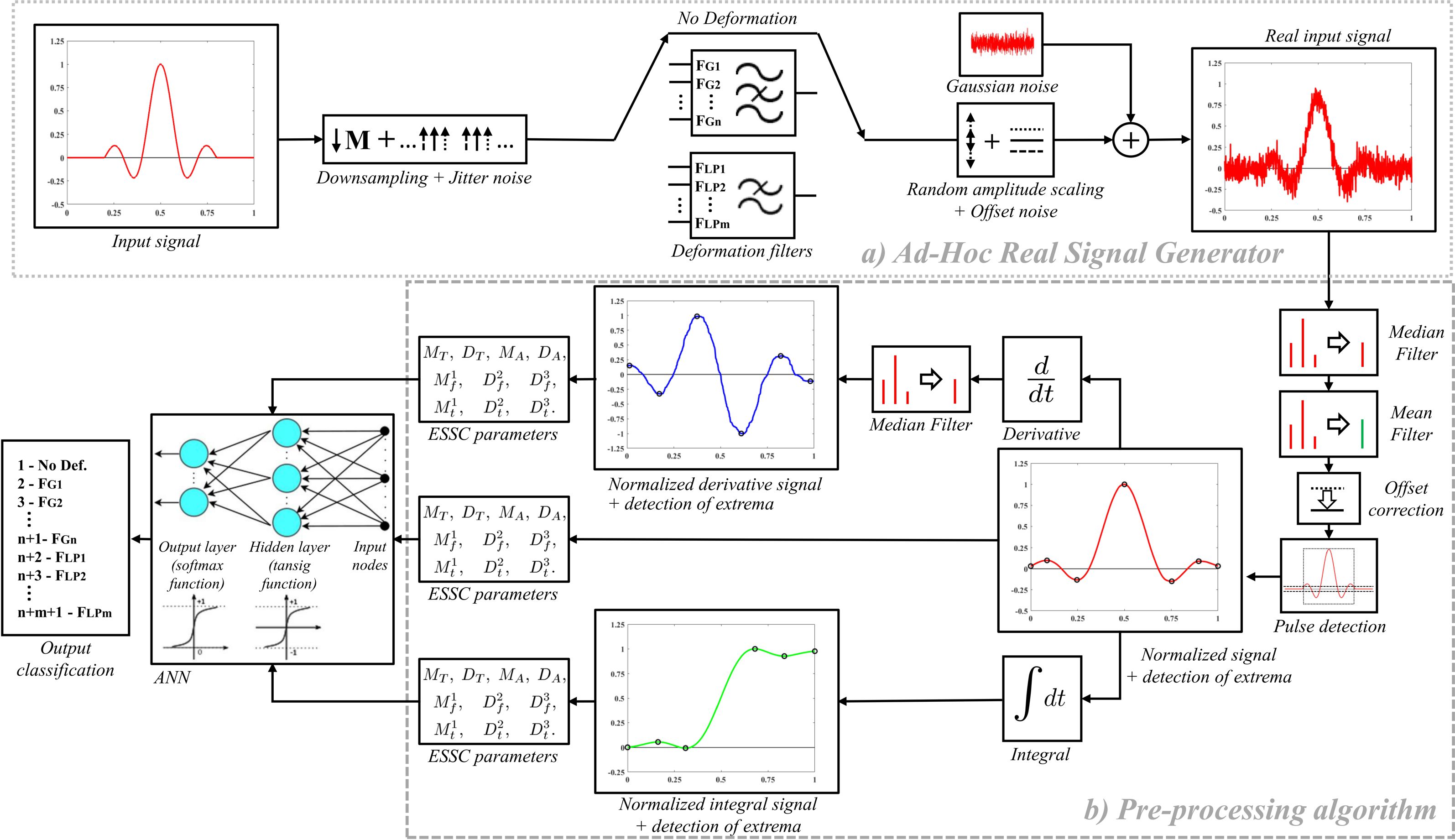}{13.5}{7.8}{Schematic of the pre-processing algorithm of the artificial neural network for the pattern-recognition method.}
    

    We propose a pre-processing algorithm whose computational strength is $\propto N_{\!s}$. Such an algorithm is detailed in \fig{fig:pre-proc}b, which illustrates the sequence of stages applied to the real input signal, enclosed within the dashed rectangle.
    In an initial step, the signal is cleaned of noise using two fast-filter stages. First, a \emph{median filter} is mainly used to clean spike noise; then, a \emph{mean filter} (or moving average filter) completes the smoothing of the function form. These moving filters are applied using windows or arrays with a few elements (around 40 for the median and 100 for the mean), which are moved over all the samples of the function. An \emph{offset correction} is applied to fix the signal's continuous level by averaging the value at each end of the signal (over a window of 100 elements).
    After these noise cleaning procedures, we perform a \emph{pulse detection} routine to extract a time and amplitude normalized form of the signal; in which the starting and ending time of the function is obtained by detecting where the absolute value of the amplitude exceeds a threshold around the 0-line (we used a 3\% of the maximum absolute value as threshold); then, these extreme values of time are reassigned as 0 and 1.\\
    The output of the pulse detection stage is the \emph{normalized signal} used to characterize the waveform, and a routine of \emph{detection of extrema} is applied to it. The obtained extrema (circles in the normalized signal waveform of \fig{fig:pre-proc}b) are used to calculate the SSC parameters of \eq{eq:SSC_Hirsch}; subsequently, we get ten ESSC parameters of the signal by calculating the first-order cumulants (see \eq{eq:stat_param}).
    The statistical characterization of the waveform is completed by calculating the \emph{derivative} and \emph{integral} of the normalized signal, and then by obtaining the \emph{ESSC parameters} from each result. Such proceedings are detailed in the branches showing the \emph{normalized derivative signal} and the \emph{normalized integral signal} waveforms, where we applied the same routine for extracting the statistical parameters as for the normalized signal (with the extrema values shown as circles in the waveforms). We note that another \emph{median filter} is applied after the derivative routine (with a window of around 20 elements); it aims to remove some spikes at the ends of the derivative function produced by the pulse detection stage, where the signal may take a step-function form due to time normalization.\\

    Finally, a 30-parameter set conforms the signal waveform identifier; it is the input to the \emph{ANN}, added as the last stage. The ANN is trained to identify the different signal deformations (i.e., the type of deformation filter used) or whether the input is a waveform with no deformation.

    In the next section, we generate datasets for training and testing ANN models.\\

\subsection{Dataset Generation for ANN training} \label{sec:data_gen}

    With the pre-processing method defined in the previous section, we now focus on building datasets to train and test ML (Machine Learning) and DL (Deep Learning) algorithms. In this way, we can compare the performance of an ML approach, utilizing various elementary ANN models and our pre-processing algorithm, with that of a DL approach. Besides, we will contrast the results obtained using the proposed 30-parameter ESSC method with those of the 4-parameter Hirsch's SSC method to quantify the improvement (or lack thereof) achieved by including the extra parameters. To carry out this task, an Ad-Hoc real signal generator \cite{GitHub_SigGen}, shown in \fig{fig:pre-proc}a within the dotted rectangle, was developed to provide the classifier models with real data.

    First, we generate an ideal pulse \emph{input signal} with a time resolution of 10000 points. Such an ideal signal passes through a \emph{downsampling} stage, where the samples are reduced by a factor of ${\rm M} = 10$ (decimation). Additionally, a uniform random time \emph{jitter noise} (between 0 and 9 samples) is applied. This stage aims to emulate the effects of an acquisition process. After that, we apply a \emph{deformation filter} ${\rm F_X}$ (where ${\rm X}$ represents low-pass, band-pass, etc.) to the signal, simulating a spectral alteration due to non-idealities in the transmission channel, or the signal passes through with \emph{no deformation}. This filter stage seeks to train the ANN with a set of known deformations (or no deformation) of the signal and then to recognize them. Following the filter stage, we perform uniform-\emph{random amplitude scaling} (up to 75\% of the maximum value), and an \emph{offset noise} introduces a uniform-random continuous level (up to 5\% of the maximum amplitude value). Then, the \emph{real input signal} is finally obtained by adding \emph{Gaussian white noise} (between $10\un{dB}$ and $25\un{dB}$), and we have a simulation of a real acquired signal with deformation (or not) to feed the pre-processing algorithm of the ANN.\\

    We are concerned with testing the feasibility of our pre-processing method of \fig{fig:pre-proc}, using ESSC parameters as inputs in a simple ANN to characterize pulse waveforms. To do that, we propose three recognition problems with the pulse form of a Sinc, a Gaussian, and a Chirp function, which are shown in (a.1) of \figs{fig:Sinc_fil}, \numfig{fig:Gauss_fil}, and \numfig{fig:Chirp_fil}, respectively. Thus, we study examples of waveforms with amplitude modulation and oscillation (Sinc), without oscillation (Gaussian), and with frequency modulation (Chirp).
    The waveforms in (a.1) have their normalized amplitude spectra shown in (a.2), where the time ($t$) and frequency ($\nu$) scales are reciprocal and arbitrary (e.g., if we have $\un{ms}$ in $t$, then we obtain $\un{kHz}$ in $\nu$, and so on).

    We chose two types of filters to apply deformations to the original pulses: one with a low-pass response and another with a sort of stop-band response that we called `Gaussian'. We name a filter with low-pass behavior as ${\rm F_{LP}}$ and a Gaussian one as ${\rm F_{G}}$ for the sake of brevity.
    The following function configures the frequency response of an ${\rm F_{LP}}$:
    \begin{equation}\label{eq:Gnu_FLP}
    G_{\rm F_{LP}}(\nu) \equiv \frac{1}{2}\left[\tanh\left({\rm SR}_\nu\,(\nu+\nu_c)\right) - \tanh\left({\rm SR}_\nu\,(\nu-\nu_c)\right)\right],
    \end{equation}
    where $\nu_c$ is the characteristic frequency and ${\rm SR}_\nu$ is the `slew rate' of the $\tanh$ profile.
    We seek to emulate the high-frequency amplitude attenuation of an amplifier or transmission channel with an ${\rm F_{LP}}$.\\
    On the other hand, the frequency response of ${\rm F_{G}}$ is configured by
    \begin{equation}\label{eq:Gnu_FG}
    G_{\rm F_{G}}(\nu) \equiv 1 - \frac{\Delta_a}{max|G_g(\nu)|}\,G_g(\nu),
    \end{equation}
    where
    \[G_g(\nu) \equiv e^{-(\nu+\nu_c)^2/(2\sigma_\nu^2)} + e^{-(\nu-\nu_c)^2/(2\sigma_\nu^2)},\]
    with $\nu_c$ and $\sigma_\nu$ being the characteristic frequency and the standard deviation of the Gaussian profile, respectively.
    In \eq{eq:Gnu_FG}, $max|G_g(\nu)|$ is the maximum absolute value of the function $G_g(\nu)$, and $\Delta_a$ is the factor of maximum attenuation (ranging between 0 and 1). The ${\rm F_{G}}$ profile emulates an irregular or no-flat amplitude response of an amplifier or transmission channel.

    To evaluate the feasibility of our pre-processing algorithm, we applied two sets of low-pass and Gaussian filters for each recognition problem involving Sinc, Gaussian, and Chirp waveforms; then, we aimed to test the neural networks' ability to differentiate between the signals and the four filtered or deformed versions associated with each waveform. The filter responses are illustrated with dotted (green) lines in \figs{fig:Sinc_fil}, \numfig{fig:Gauss_fil}, and \numfig{fig:Chirp_fil}, specifically in panels (b.2, c.2, d.2, and e.2), where we also indicate the parameters of the functions defined in \eqs{eq:Gnu_FLP} and \numeq{eq:Gnu_FG}.
    The deformed signal is generated by multiplying the original amplitude spectrum (a.2) by the filter response, then applying the inverse Fourier transform to the resulting product. In our study, we considered the no-deformed signal (a.1 - ${\rm ND}$) along with its amplitude spectrum (a.2), and the deformed signals (b.1 - ${\rm F_{G1}}$, c.1 - ${\rm F_{G2}}$, d.1 - ${\rm F_{LP1}}$, and e.1 - ${\rm F_{LP2}}$), with their spectra displayed as solid (blue) lines in panels (b.2, c.2, d.2, and e.2).

    We propose these basic recognition pattern examples as an initial test for our algorithm's performance, encompassing both lightly and heavily deformed signals. To simulate real measurements, we used the algorithm illustrated in \fig{fig:pre-proc}a to generate a set of random-noisy waveforms that reflect the specified filter deformations: ${\rm ND}$, ${\rm F_{G1}}$, ${\rm F_{G2}}$, ${\rm F_{LP1}}$, and ${\rm F_{LP2}}$. From these waveforms, we compiled datasets that include their ESSC and SSC parameters for training ML models \cite{GitHub_MLdata}. Notably, the SSC dataset is constructed by extracting the four parameters shown in \eq{eq:SSC_Hirsch} from the 30-parameter ESSC dataset. Additionally, we collected raw waveforms containing 256 and 1024 samples to train DL networks \cite{GitHub_DLdata}. Each dataset was generated at different signal-to-noise ratios (SNR), ranging from $25\un{dB}$ to $10\un{dB}$ in $5\un{dB}$ increments. \fig{fig:Fig_Dataset_List} illustrates these datasets, and we will discuss the results of the proposed recognition problems in the following sections.
    We have included in refs. \cite{GitHub_SigGen, GitHub_MLdata, GitHub_DLdata} waveform samples generated by the Ad-Hoc real signal generator at varying SNR levels, illustrating the increasing difficulty of the pattern recognition problem as noise levels rise.

    In the next section, we define the type of ANN to be implemented in the pre-processing algorithm among the different models available in traditional ML and DL environments. The selection criterion is based on model simplicity, accuracy, and performance, as well as training and classification time, for optimization.

	\insertfig{h}{fig:Sinc_fil}{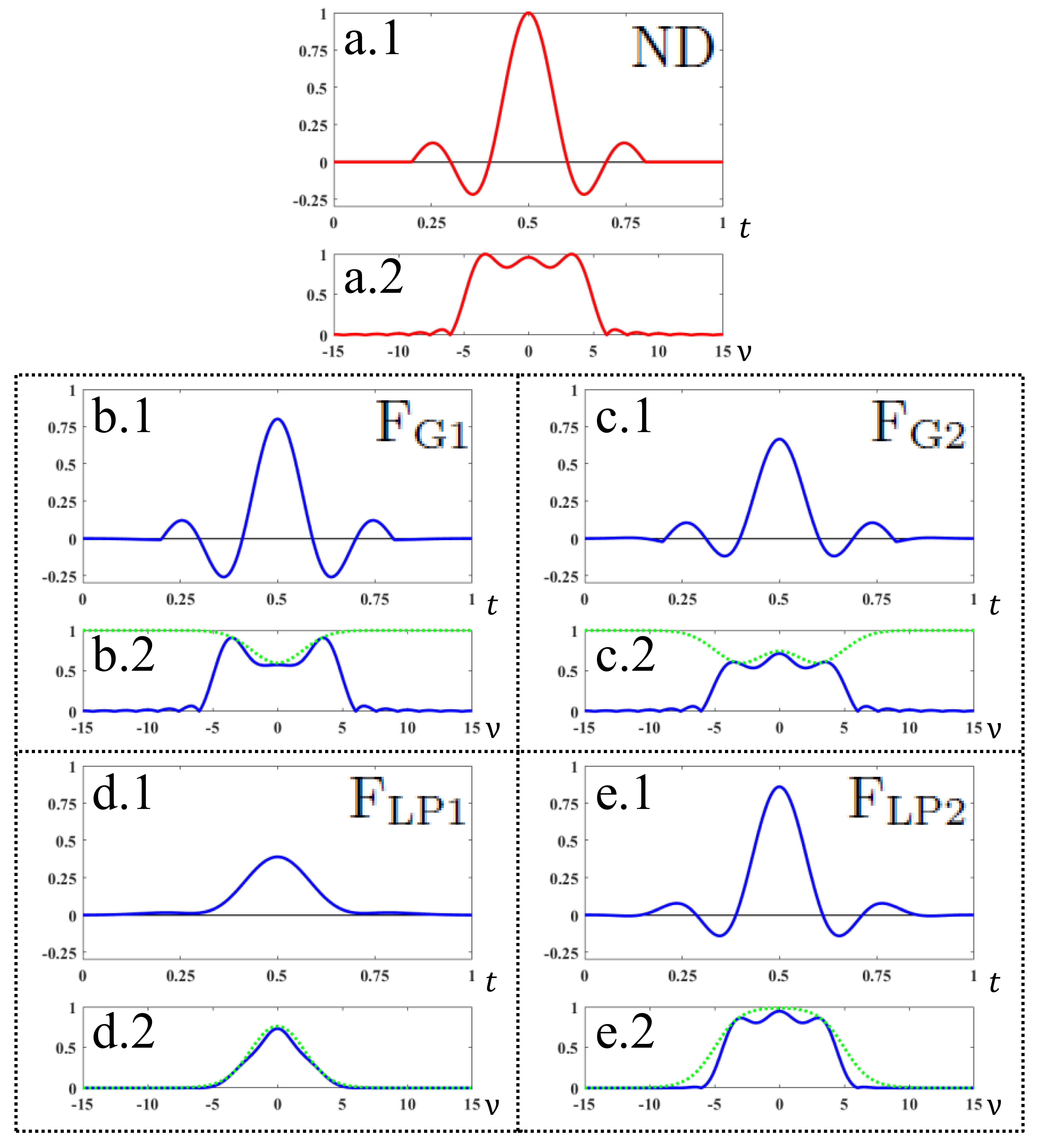}{11.5}{12.7}{Deformation filters applied to a Sinc pulse. The original Sinc signal (a.1, no deformation) and its amplitude spectrum (a.2). It is shown the resulting signals (b.1, c.1, d.1, and e.1) and their corresponding amplitude spectra (solid blue line in b.2, c.2, d.2, and e.2) obtained by multiplying the original signal spectrum (a.2) times the deformation filter amplitude spectra (dotted green line in b.2, c.2, d.2, and e.2). The parameters for Gaussian and Low-Pass filters are the following, (b.2) ${\rm F_{G1}}$: $\nu_c = 0$, $\sigma_\nu = 2$, $\Delta_a = 0.4$; (c.2) ${\rm F_{G2}}$: $\nu_c = 3$, $\sigma_\nu = 2$, $\Delta_a = 0.4$; (d.2) ${\rm F_{LP1}}$: $\nu_c = 2$, ${\rm SR}_\nu = 0.5$; (e.2) ${\rm F_{LP2}}$: $\nu_c = 5$, ${\rm SR}_\nu = 0.5$.}
	
	\insertfig{h}{fig:Gauss_fil}{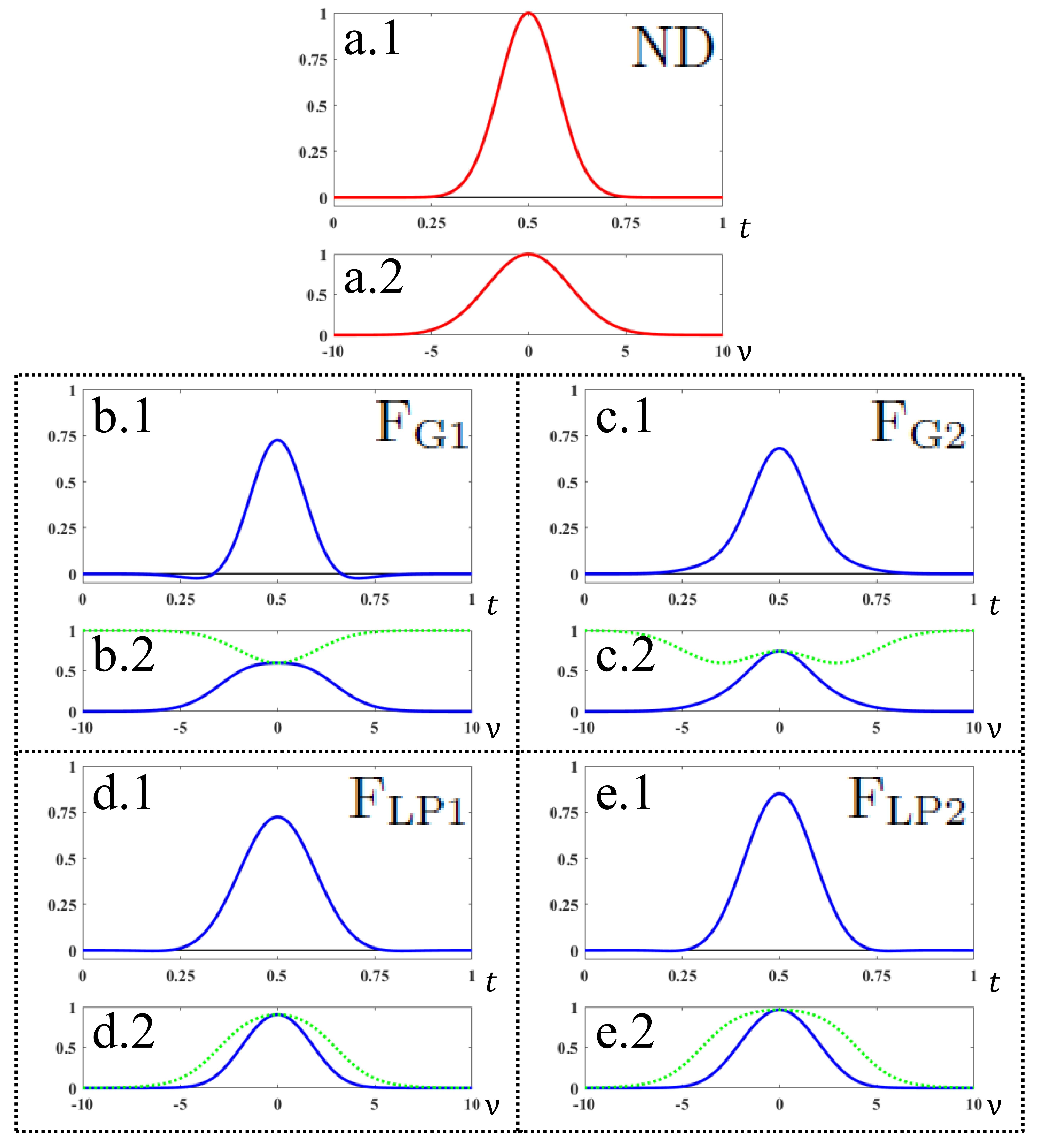}{11.5}{12.7}{Deformation filters applied to a Gaussian pulse. The original Gaussian signal (a.1, no deformation) and its amplitude spectrum (a.2). It is shown the resulting signals (b.1, c.1, d.1, and e.1) and their corresponding amplitude spectra (solid blue line in b.2, c.2, d.2, and e.2) obtained by multiplying the original signal spectrum (a.2) times the deformation filter amplitude spectra (dotted green line in b.2, c.2, d.2, and e.2). The parameters for Gaussian and Low-Pass filters are the following, (b.2) ${\rm F_{G1}}$: $\nu_c = 0$, $\sigma_\nu = 2$, $\Delta_a = 0.4$; (c.2) ${\rm F_{G2}}$: $\nu_c = 3$, $\sigma_\nu = 2$, $\Delta_a = 0.4$; (d.2) ${\rm F_{LP1}}$: $\nu_c = 3$, ${\rm SR}_\nu = 0.5$; (e.2) ${\rm F_{LP2}}$: $\nu_c = 4$, ${\rm SR}_\nu = 0.5$.}
	
	\insertfig{h}{fig:Chirp_fil}{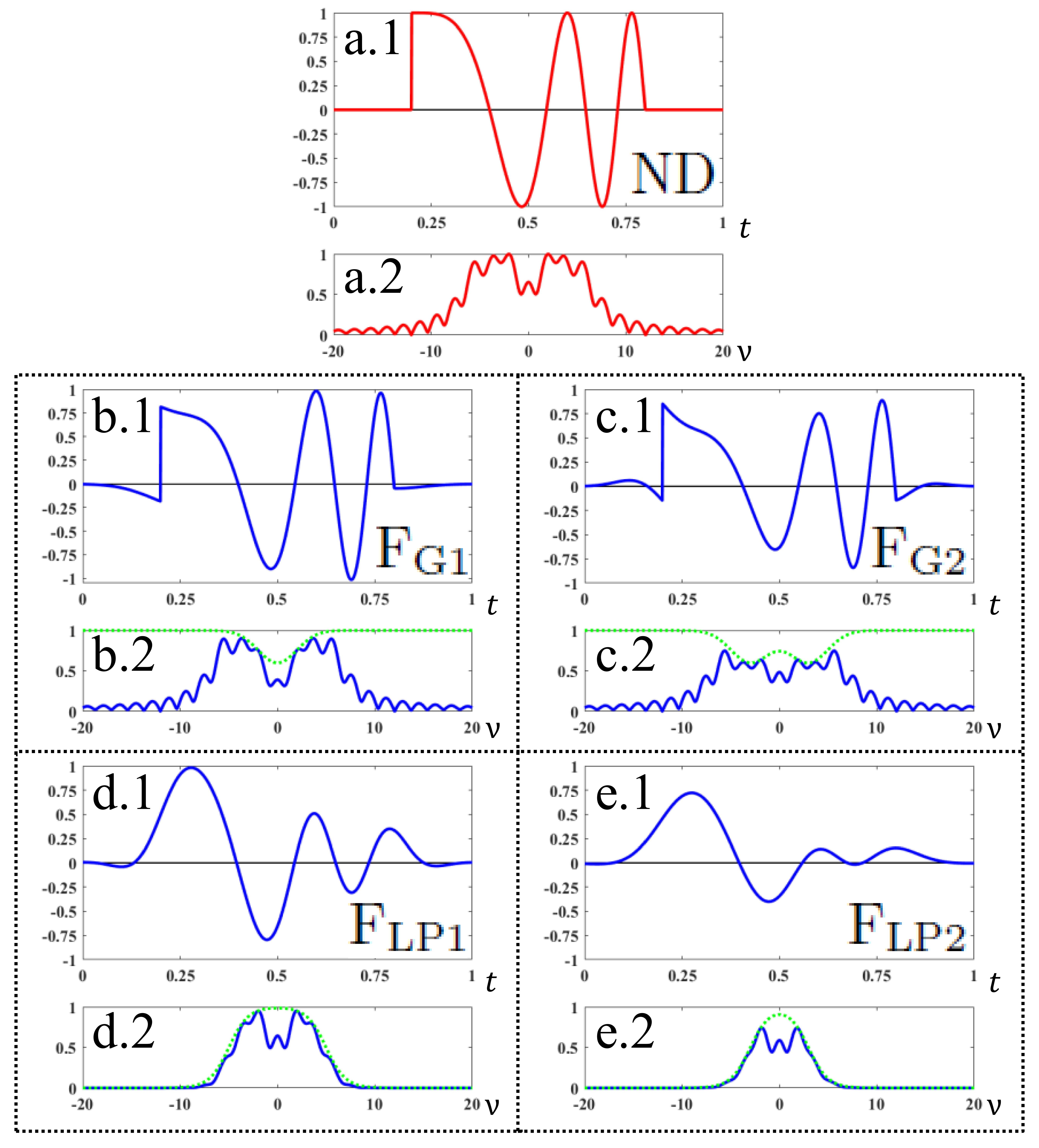}{11.5}{12.7}{Deformation filters applied to a Chirp pulse. The original Chirp signal (a.1, no deformation) and its amplitude spectrum (a.2). It is shown the resulting signals (b.1, c.1, d.1, and e.1) and their corresponding amplitude spectra (solid blue line in b.2, c.2, d.2, and e.2) obtained by multiplying the original signal amplitude (a.2) times the deformation filter amplitude spectra (dotted green line in b.2, c.2, d.2, and e.2). The parameters for Gaussian and Low-Pass filters are the following, (b.2) ${\rm F_{G1}}$: $\nu_c = 0$, $\sigma_\nu = 2$, $\Delta_a = 0.4$; (c.2) ${\rm F_{G2}}$: $\nu_c = 3$, $\sigma_\nu = 2$, $\Delta_a = 0.4$; (d.2) ${\rm F_{LP1}}$: $\nu_c = 5$, ${\rm SR}_\nu = 0.5$; (e.2) ${\rm F_{LP2}}$: $\nu_c = 3$, ${\rm SR}_\nu = 0.5$.}

	\insertfig{h!}{fig:Fig_Dataset_List}{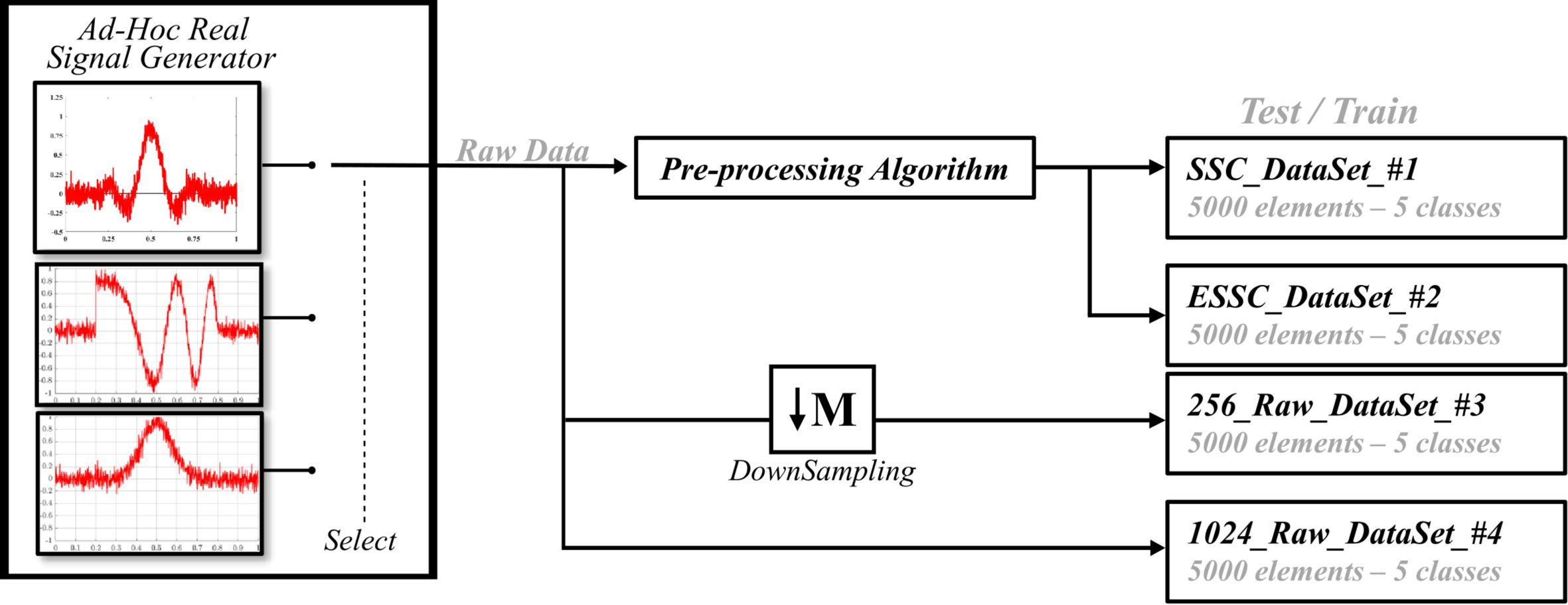}{10.0}{3.86}{Balanced Dataset Scheme generated to train and test ML and DL algorithms with a total of 5000 elements, 1000 per class.}
	

\subsection{ESSC and raw-waveform data performance measurement in traditional supervised Machine Learning classifiers and Deep Learning models} \label{sec:param_perf}

    With the balanced training and test datasets established in the previous section, we aim to evaluate the performance of using ESSC parameters with traditional supervised Machine Learning (ML) classifiers under varying Signal-to-Noise Ratio (SNR) conditions for pulsed waveforms. Additionally, we will compare the results with those obtained from Deep Learning (DL) algorithms.  Various ML classification models can effectively tackle pattern recognition problems when suitable training methods, parameter tuning, and datasets are used for both training and testing.
    In \appen{app:descMLDL}, we provide a brief overview of the most common ML methods, including Naive Bayes (NB), K-Nearest Neighbors (KNN), Support Vector Machine (SVM), and Multi-Layer Perceptron (MLP), highlighting the main features of each method. Besides, we discuss Convolutional Neural Networks (CNN) for DL applications. This information allows us to establish criteria for selecting the most appropriate method for the recognition task.

    To apply each ML and DL model to practical problems effectively requires optimal parameter tuning. Parameters can be classified into two categories: those related to the model architecture and must be set before the learning process (hyperparameters), and those that can be updated during the learning process (Model Parameters) \cite{ Li_Yang_2020}.
    When working with highly flexible models like MLPs and CNNs, the primary hyperparameters include the number of hidden layers, the number of neurons, the choice of loss function, the activation function, learning rates, and regularization techniques. In the case of SVM, important parameters are kernel selection and the C regularization parameter. Finding optimal values for these parameters is essential for creating an optimized model and ensuring a fairer comparison of classifier performance.
    At this stage, manually configuring parameters is both impractical and prone to suboptimal model tuning; moreover, it can lead to issues, such as getting stuck in a local minimum, when using models like MLPs and CNNs while minimizing the loss function. For this research, we used Hyperparameter Bayesian Optimization (HBO) techniques with the MATLAB{\circledR} Toolbox, performing 30 training iterations on the hardware configuration specified in \tabla{tab:sys_resources}.

    After training, we conducted recognition pattern experiments, as detailed in \sect{sec:data_gen}, for Sinc, Gaussian, and Chirp waveforms. We used the specified ML and CNN models and optimized them using the mentioned toolbox. The results obtained for each model are similar to those presented in the upcoming section for the MLP case, where a Confusion Matrix (CM) illustrates the number of successful recognitions for each model when identifying the different signal waveforms. The accuracy of each ML model is defined as the ratio of successful outputs to the total number of inputs, expressed as a percentage (calculated by summing the diagonal values of the CM divided by the sum of all values in the matrix). The accuracy performance results are presented in \fig{fig:bar_plot_Acc} for Sinc, Gaussian, and Chirp signals, under different SNR conditions. Moreover, we measured the computation time required to classify a single sample signal, using ESSC parameters generated by the proposed pre-processing technique. To accomplish this, we used the \emph{timeit} function (see the MATLAB{\circledR} documentation for more details). The measurement results, with an SNR of $25\un{dB}$, are presented in \fig{fig:bar_plot_time}, where each processing time value represents the average of 100 algorithm executions. In \tabla{tab:train_time_mldl}, we present the training times for each network using ML and DL algorithms, as reported by the toolbox upon completion of execution, serving as a measure of the computational effort required to develop each architecture that addresses the proposed pattern recognition problem.

    We observe that the tested ML methods have similar accuracy at the same SNR level, with values near 90\% at SNR levels above $20\un{dB}$ in at least one method (see \fig{fig:bar_plot_Acc}), as is the case with the 1D-CNN algorithm. However, at lower SNR values, overall performance is significantly lower with ML algorithms than with 1D-CNN. On the other hand, ML techniques that utilize the ESSC 30-parameter dataset consistently outperform those that use the SSC 4-parameter dataset in nearly all cases. Notably, the performance with the ESSC dataset shows significant improvement for a Gaussian waveform (see \figs{fig:bar_plot_Acc} (a.2) and (b.2)), which can be attributed to the very low number of extrema present in the Gaussian form.
    As shown in \fig{fig:bar_plot_time}, the KNN, SVM, and MLP algorithms are faster than the 1D-CNN algorithms in classifying a single data register from the dataset. Remarkably, the MLP network is the fastest option, operating approximately 4 times faster than the 1D-CNN with 256 elements.

    Finally, comparing the training times in \tabla{tab:train_time_mldl} shows that constructing DL networks, even with just 256 elements, demands significantly more complexity and computational effort than building ML networks.
    In conclusion, given the comparable accuracy of all the ML networks in pattern recognition tasks and the MLP's fastest classification speed, we selected the MLP as the ANN for a detailed evaluation of the proposed pre-processing algorithm's performance. We will elaborate on this in the next section.

	\begin{table}[h!]
	\setlength\tabcolsep{0pt} 
	\caption{System resources for training and testing.}
	\label{tab:sys_resources}
	\begin{tabular*}{\textwidth}{@{\extracolsep{\fill}} lc}
	\toprule
	   Resources & Details  \\
	\midrule
	    Processor           & Intel(R) Core(TM) Ultra 7 155U  @ 1.70 GHz \\
	    System Type         &  x64-based processor  \\
	    RAM                 &  32.0 GB @ 5600Mhz  \\
	    Operation Systems   &  Windows 11 Home  \\
	\bottomrule
	\end{tabular*}
	\end{table}

	\insertfig{hp}{fig:bar_plot_Acc}{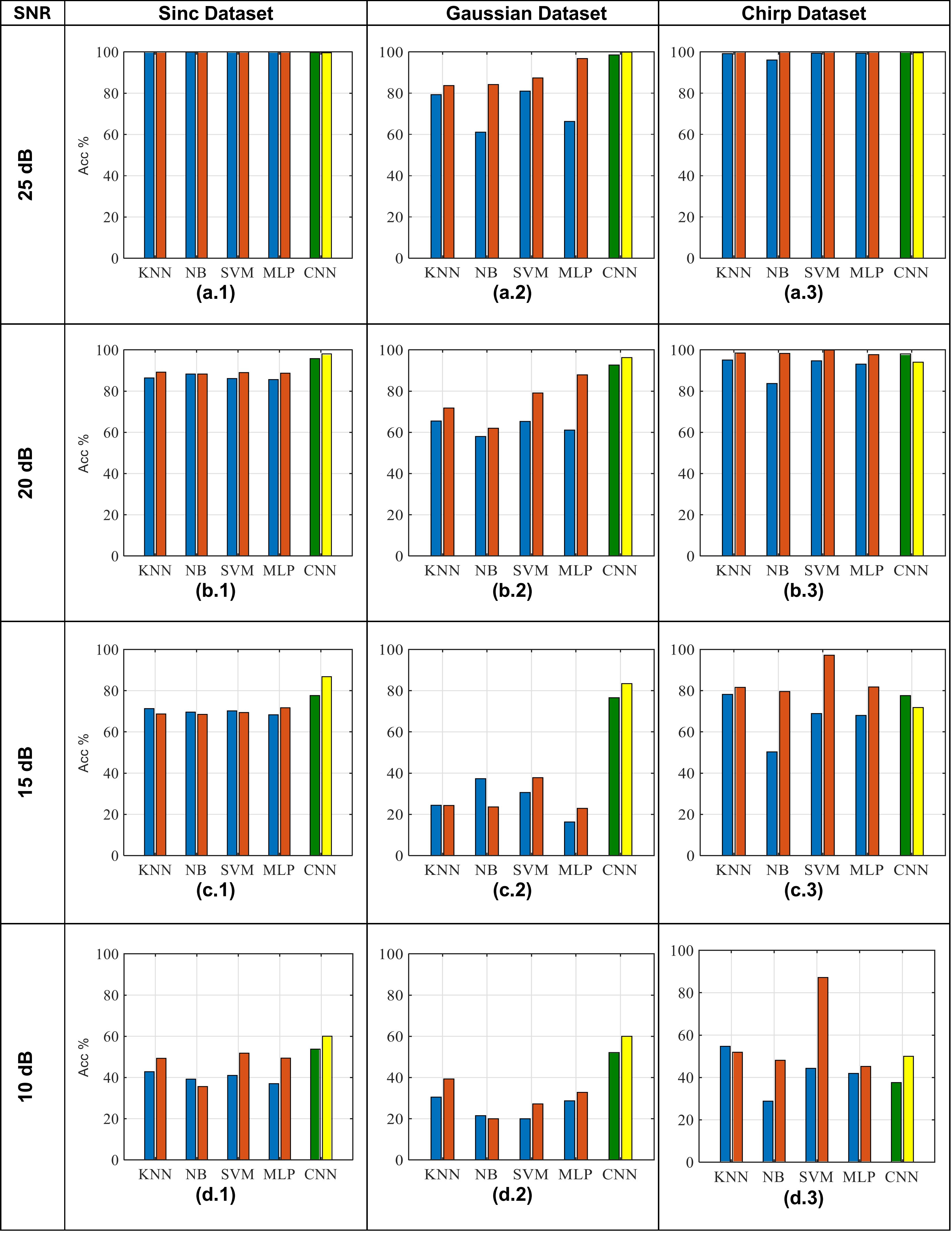}{14.0}{18.1}{Accuracy measurement (in \%) for Machine (KNN, NB, SVM, and MLP) and Deep Learning (1D-CNN) algorithms at different SNR levels for Sinc, Gaussian, and Chirp waveforms. Where SSC (blue bars) and ESSC (red bars) parameters (4 and 30 elements, respectively) for ML, and samples with 256 (green bars) and 1024 (yellow bars) elements for DL, are used as the input dataset.}
	

	\insertfig{hp}{fig:bar_plot_time}{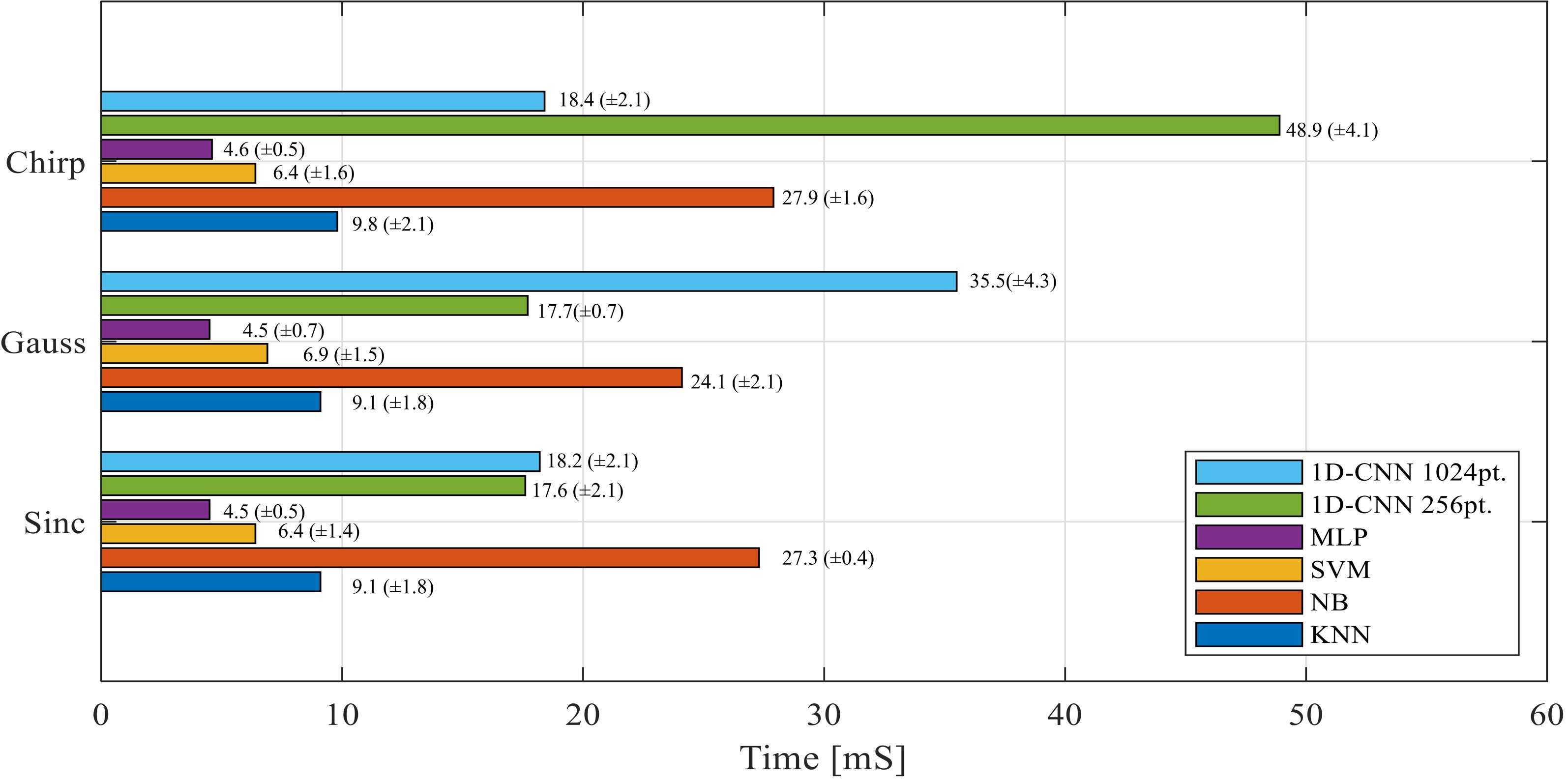}{13.5}{6.72}{Average computation time to classify one data register for Sinc, Gaussian, and Chirp waveforms (using 100 algorithm executions) with Gaussian noise at a $25\un{dB}$ SNR level, utilizing an ANN model trained with ESSC (30 parameters) for Machine Learning (KNN, NB, SVM, and MLP), and with 256 and 1024 sample elements for Deep Learning (1D-CNN). The time is measured using the \emph{timeit} MATLAB{\circledR} function.}
	
	
	\begin{table}[h!]
	\centering
	\setlength\tabcolsep{0pt} 
	\caption{Training time for Machine and Deep Learning models under Hyperparameter Bayesian Optimization expressed in seconds for different waveform recognition pattern problems (Sinc, Gaussian, and Chirp), using SSC and ESSC parameters (4 and 30 elements, respectively)  for ML, and  256 and 1024 sample elements for DL. The MATLAB{\circledR} Toolbox supplies such a time upon completion of its execution.}
	\label{tab:train_time_mldl}
	\begin{tabular*}{\textwidth}{@{\extracolsep{\fill}} lccccccc}
	\toprule
	        & \multicolumn{2}{c}{Sinc} & \multicolumn{2}{c}{Gaussian} & \multicolumn{2}{c}{Chirp} \\
	        \cmidrule(r){2-3}  \cmidrule(r){4-5}  \cmidrule(r){6-7}
	Parameter &  SSC   &     ESSC &    SSC  &    ESSC &    SSC  &    ESSC \\
	        \cmidrule(r){2-3}  \cmidrule(r){4-5}  \cmidrule(r){6-7}
	    KNN & 42.017     & 43.6        & 34.913      & 29.663      & 22.269      & 73.022      \\
	    NB  & 84.972     & 400.36      & 100.39      & 423.34      & 68.811      & 344.07      \\
	    SVM & 113.25     & 135.45      & 584.82      & 145.17      & 128.46      & 203.82      \\
	    MLP & 120        & 560.65      & 1542.5      & 1510.8      & 2031.8      & 736.35      \\
	\midrule
	Sample size &  256   &  1024 &   256   &  1024 &    256   &  1024 \\
	        \cmidrule(r){2-3}  \cmidrule(r){4-5}  \cmidrule(r){6-7}
	    1D-CNN & 3601.12  & 13505.04  & 2722.64 & 13996.6 & 3109.39 & 15641.35      \\
	\bottomrule
	\end{tabular*}
	\end{table}

\subsection{Results for the pattern recognition problem using the ESSC pre-processing method with an MLP network}\label{sec:res_ex-pr}

    In this section, we analyze the results of the pattern recognition problem presented in \sect{sec:data_gen} using the MLP network. We aim to thoroughly test the performance of our proposed pre-processing algorithm using the ESSC 30-parameter feature extraction method introduced in \sects{sec:def_ESSC} and \numsect{sec:preproc_algorithm}. Accordingly, we apply the algorithm of \fig{fig:pre-proc} to recognize several pulse waveforms affected by different spectral deformations (see \figs{fig:Sinc_fil}, \numfig{fig:Gauss_fil}, and \numfig{fig:Chirp_fil}).

    The MLP structure (see the \emph{ANN} block in \fig{fig:pre-proc}) is shown in \tabla{tab:mlp_params}, where the input layer has 30 nodes; this number corresponds to the ESSC parameters entries. Then, the hidden layers and activation function were defined using the Hyperparameter Bayesian Optimization (HBO) training method, as mentioned in \sect{sec:param_perf}. Finally, the output layer includes neurons and output terminals corresponding to the number of classes to be predicted; in our case, it is 5 (${\rm ND}$, ${\rm F_{G1}}$, ${\rm F_{G2}}$, ${\rm F_{LP1}}$, and ${\rm F_{LP2}}$). These output neurons utilize the \emph{softmax} activation function and employ the \emph{one-hot encoding method}. The softmax's output is a 5-element array containing the predicted probability for each class, ranging from 0 to 1.

    We executed the algorithm in \fig{fig:pre-proc} for each output class and generated a set of 1000 elements, each containing a 30-parameter array (5000 elements in total), for the training process. Additionally, we utilize a Gaussian White Noise (GWN) source with a zero mean and a standard deviation ($\sigma_{\opc{N}}$) set as 5\% of the maximum signal amplitude ($\abs{f}_{max}$). Then, we obtain an ANN for each recognition problem (Sinc, Gaussian, and Chirp).

    After constructing the Artificial Neural Network (ANN), we utilize it to evaluate the effectiveness of the pre-processing algorithm in our proposed examples.
    In our instances, the classification problems involve different \emph{classes}, each identified by a specific ID number: ${\rm ND}$ (1) represents a non-deformed signal, while ${\rm F_{G1}}$ (2), ${\rm F_{G2}}$ (3), ${\rm F_{LP1}}$ (4), and ${\rm F_{LP2}}$ (5) correspond to various deformations caused by filtering

    The ANN output is a number that identifies the input class type. A successful class identification occurs when the predicted output class matches the actual input class. Therefore, we can test the network's ability to recognize a class by knowing the target class of the input and comparing it with the resulting output. Accordingly, the performance of a pattern-recognition exercise is reflected by using a \emph{confusion matrix}, which indicates the number of times that an \emph{output class} is obtained due to a fixed \emph{target class}.\\
    We conduct testing for the ESSC method across various GWN levels, with an SNR value ranging from $25\un{dB}$ to $10\un{dB}$ in $5\un{dB}$ increments, where ${\rm SNR} = 20\log_{10}\left(\abs{f}_{max}/\sigma_{\opc{N}}\right)$ (approximately between a 5.6\% and 31.6\% ratio of $\sigma_{\opc{N}}/\abs{f}_{max}$). In this way, we used a dataset of 5000 elements or statistical parameters (1000 per class) for each pulse waveform and noise level (similar to the training dataset). The results are presented in \figs{fig:CM_Sinc}, \numfig{fig:CM_Gauss}, and \numfig{fig:CM_Chirp} for Sinc, Gaussian, and Chirp signals, and they are analyzed in \sect{sec:disc_conc}.

    To conclude this section, we study the relevance of each parameter using the ReliefF algorithm \cite{Urbanowicz2018}.
    We use the MATLAB{\circledR} ReliefF function for classification, and the results for the recognition examples (with an SNR of $25\un{dB}$) are presented in \fig{fig:ReliefF}, where $K=10$ is set as the parameter for the number of nearest neighbors per class.
    We create a graph for each pulse waveform, displaying a bar for the signal (middle red bar), the derivative (left green bar), and the integral (right blue bar). Therefore, there are ten parameter names on the abscissa, and three values are assigned to each (corresponding to the ESSC 30 parameters). A parameter is relevant if its value exceeds the threshold $\tau$, which can be defined using Chebyshev's inequality, as is known.
    Then, for a given confidence level $\alpha$, $\tau = 1/\sqrt{\alpha\,m}$ is good enough to make the probability of a Type I error less than $\alpha$, where $m$ is the number of random training instances used to define the ReliefF values. We use all the dataset values; thus, $m = 5000$.
    The higher values of ReliefF define the higher bound for $\tau$ (and the lower bound of $\alpha$); then, approximately
    $\tau = 0.41,\, 0.17,\,\text{and}\, 0.46$ (or $\alpha = 0.0012,\, 0.0069,\,\text{and}\, 0.0009$) is obtained for the Sinc, Gaussian, and Chirp examples, respectively.
    Thus, the best-obtained condition limits the error to approximately $0.1\%$. Besides, all the ReliefF values are positive.
    We conclude that the 30 parameters are relevant in the proposed recognition or classification examples, with varying degrees of participation depending on the pattern to be recognized or characterized. In particular, the lower values obtained for the Gaussian problem reflect the difficulty of this signal discrimination task (see \fig{fig:Gauss_fil}).

    \begin{table}[h!]
        \centering
        \setlength\tabcolsep{0pt} 
        \caption{Multi-Layer Perceptron parameters configuration for Sinc, Gaussian, and Chirp signals after Hyperparameter Bayesian Optimization (HBO).}
        \label{tab:mlp_params}
        \begin{tabular*}{\textwidth}{@{\extracolsep{\fill}} lcccc}
        \toprule
            --      & Sinc MLP   & Gaussian MLP    & Chirp MLP \\
        \midrule
            \textbf{Hyperparameter}         &           &           &                   \\
            Optimization Technique          & HBO       &    HBO    &   HBO                \\
            Iterations                      & 30        &    30     &   30                \\
            Data Standardization            & No        &    Yes    &   Yes             \\
            Number of fully connected layers   & 2         & 2         & 3                 \\
            Activation Function             & Tangh     & None      & ReLu              \\
            Lambda                          & 7.918e-08 & 8.6864e-05& 0.0002154         \\
            First layer size                & 8         & 14        & 2                 \\
            Second layer size               & 3         & 278       & 136               \\
            Third layer size                & -         & -         & 285               \\
        \midrule
            Input Params ESSC num           & 30        & 30        & 30               \\
            Output classes num              & 5         & 5         & 5               \\
        \bottomrule
        \end{tabular*}
    \end{table}

	\insertfig{hp}{fig:CM_Sinc}{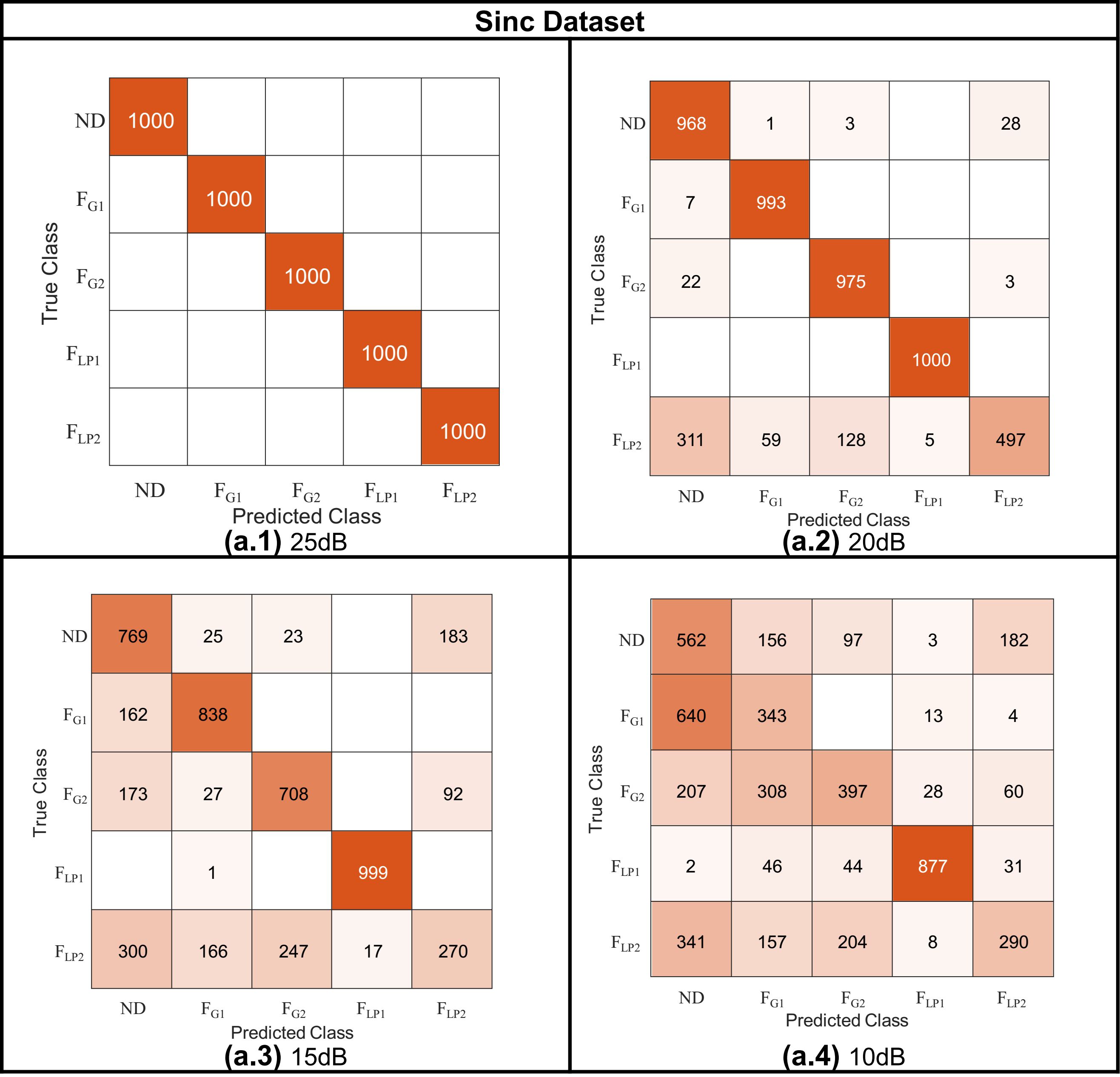}{11.0}{10.8}{Confusion matrix for the pattern-recognition example with a Sinc-pulse signal. The results are presented for various SNR levels of GWN, using the ESSC pre-processing algorithm with a Multi-Layer Perceptron. The network was trained using a Hyperparameter Bayesian Optimization method and tested on a dataset of 5000 elements (1000 per class), each an array of 30 ESSC parameters.}
	

	\insertfig{hp}{fig:CM_Gauss}{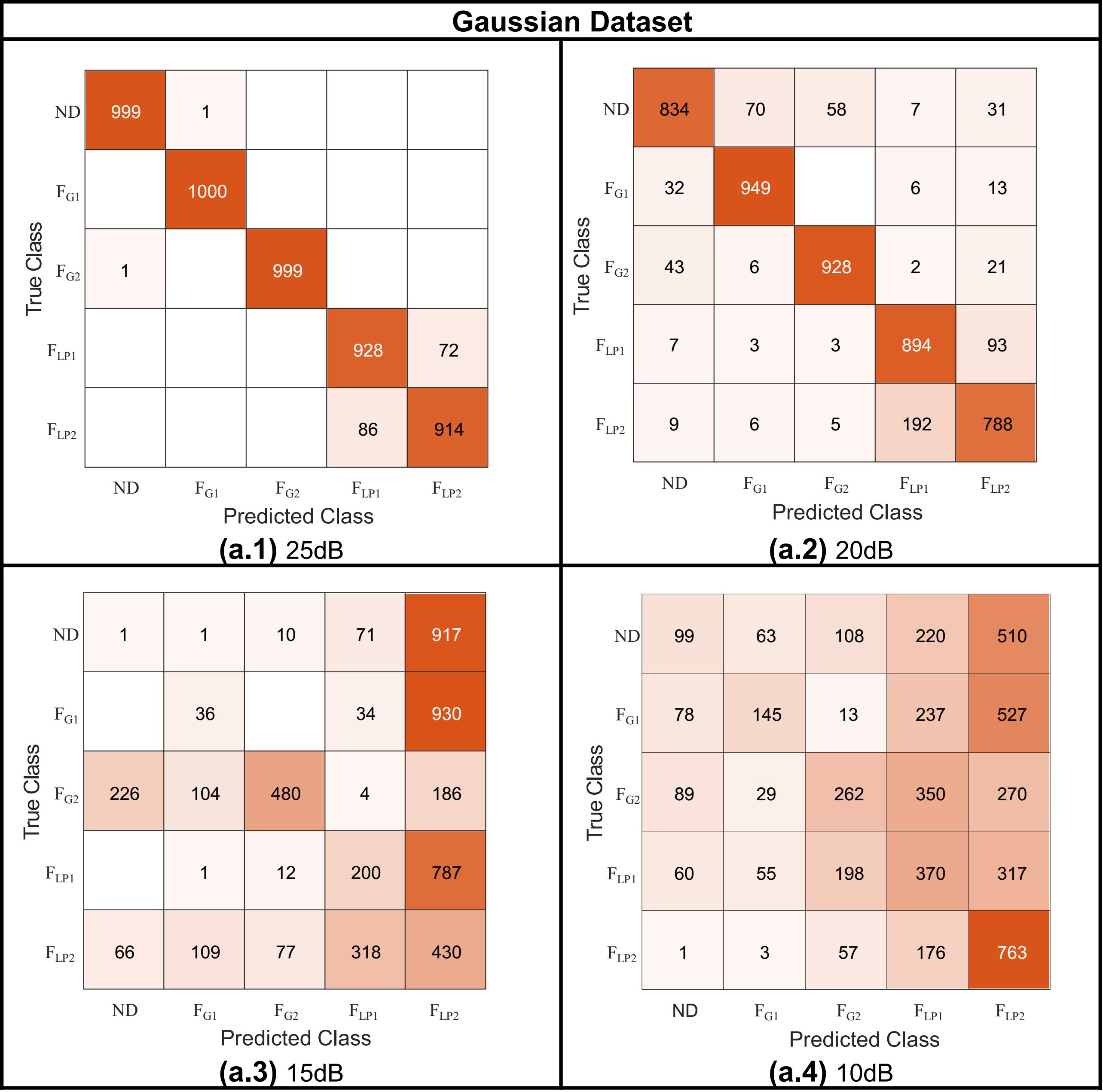}{11.0}{10.8}{Confusion matrix for the pattern-recognition example with a Gaussian-pulse signal. The results are presented for various SNR levels of GWN, using the ESSC pre-processing algorithm with a Multi-Layer Perceptron. The network was trained using a Hyperparameter Bayesian Optimization method and tested on a dataset of 5000 elements (1000 per class), each an array of 30 ESSC parameters.}
	

	\insertfig{hp}{fig:CM_Chirp}{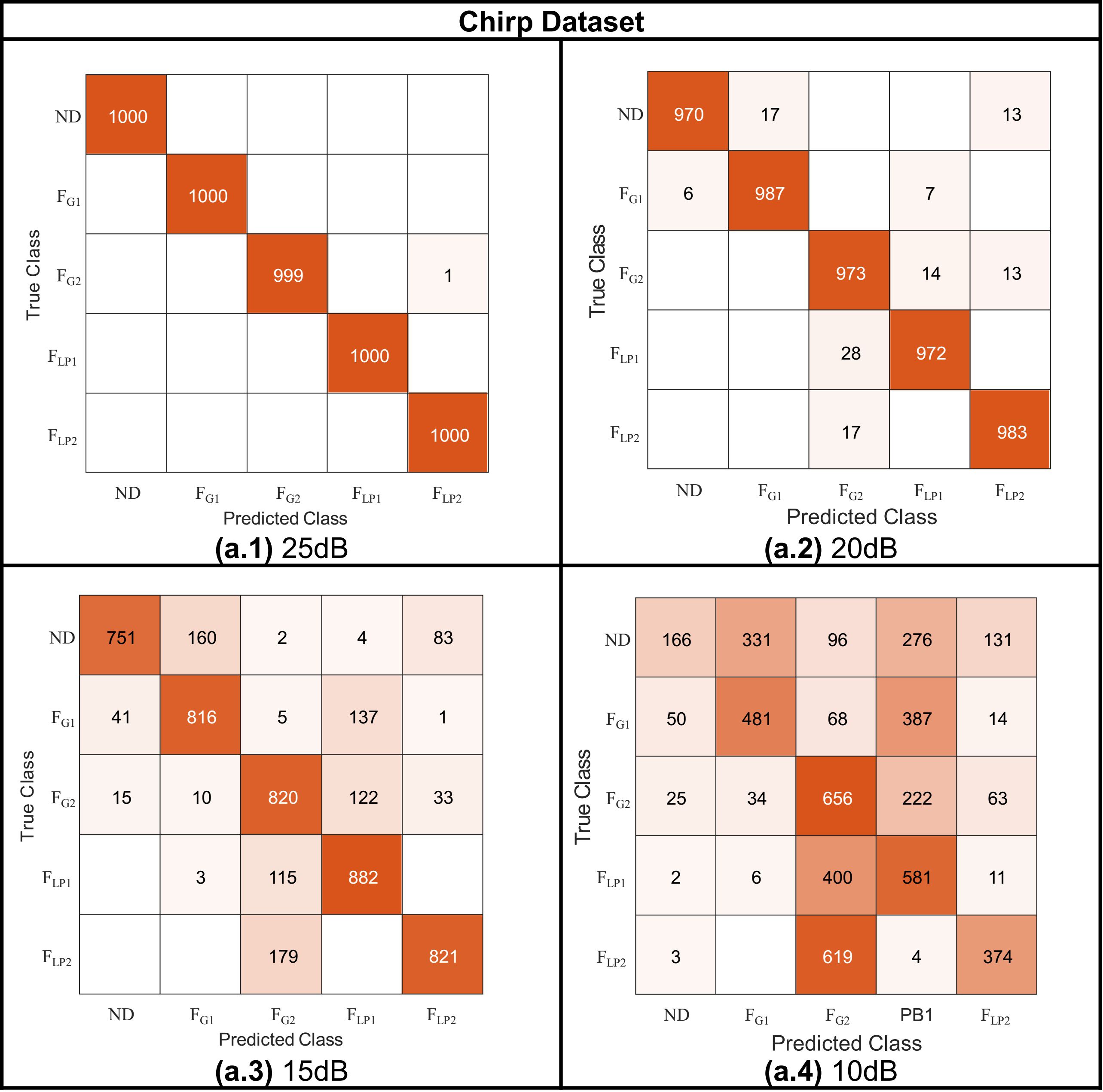}{11.0}{10.8}{Confusion matrix for the pattern-recognition example with a Chirp-pulse signal. The results are presented for various SNR levels of GWN, using the ESSC pre-processing algorithm with a Multi-Layer Perceptron. The network was trained using a Hyperparameter Bayesian Optimization method and tested on a dataset of 5000 elements (1000 per class), each an array of 30 ESSC parameters.}
	

	\insertfig{hp}{fig:ReliefF}{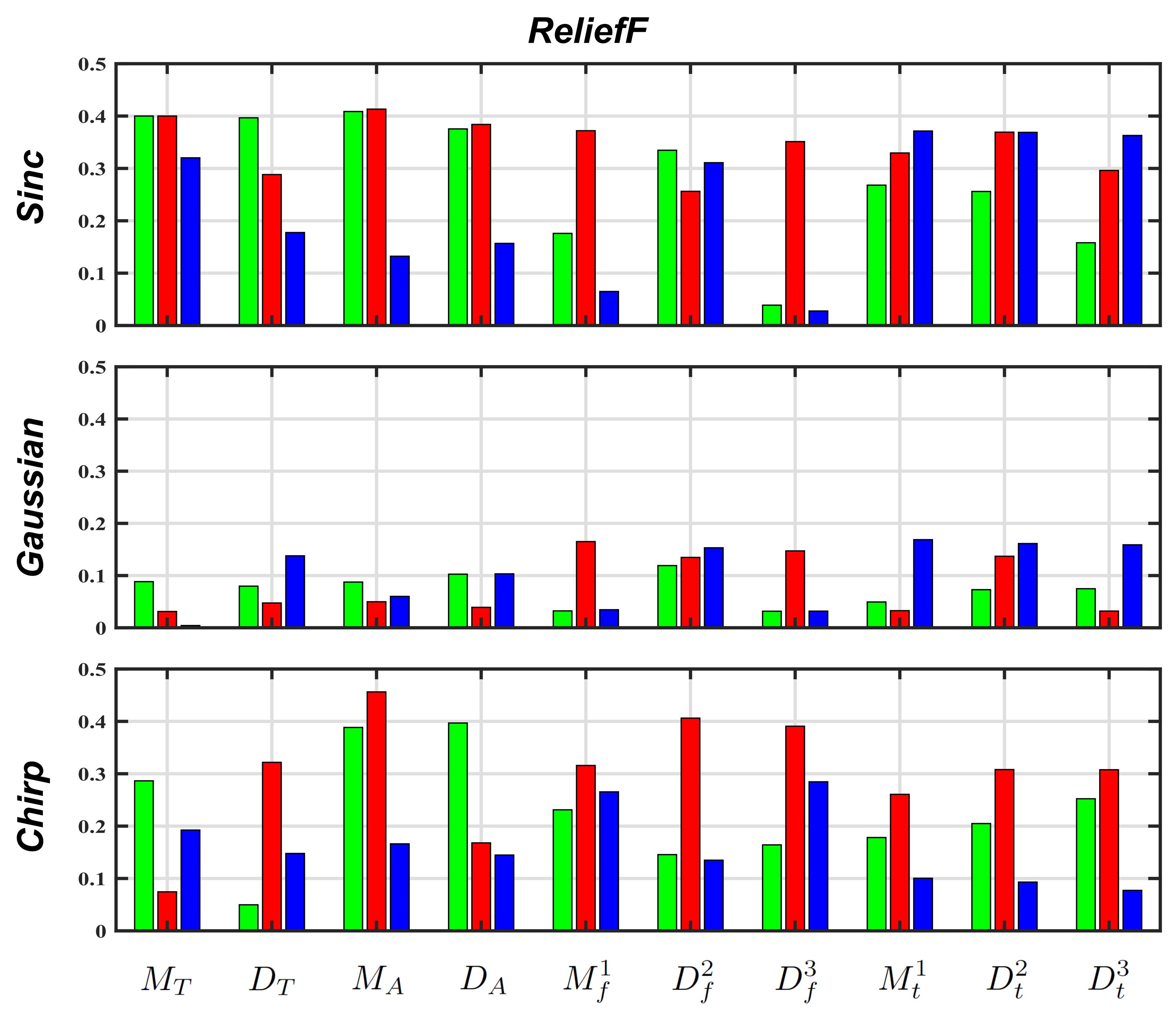}{13.5}{11.8}{MATLAB{\circledR} ReliefF function values of each ESSC parameter for the Sinc, Gaussian, and Chirp pattern recognition example, for classification with the parameter $K=10$. Middle red bar: signal, left green bar: derivative, right blue bar: integral.}


\section{Discussion and conclusion}\label{sec:disc_conc}

    In this work, we propose a 30-parameter pre-processing method to extract waveform features that can be applied in a pattern recognition task (\sects{sec:mathapp} and \numsect{sec:app_ESSC_ANN}) employing traditional Machine Learning (ML) algorithms. Such a method is based on the statistical characterization of the signal, using moments and cumulants of the waveform, its derivative, and its integral (\sect{sec:def_ESSC}). The proposed technique is presented as an extension of the SSC method developed by H.L. Hirsch \cite{Book_Hirsch92,Hirsch92}, and therefore, we refer to it as ESSC.\\
    When analyzing a waveform consisting of $N_{\!s}$ samples, calculating statistical parameters involves summation operations that scale linearly with $N_{\!s}$. In this process, we extract features such as the mean, variance, skewness, and kurtosis. In contrast, feature extraction using the Fast Fourier Transform (FFT) or convolution entails a more complex set of operations, scaling as $N_{\!s} \log_2(N_{\!s})$. Consequently, the ESSC technique requires minimal computational effort because it relies on statistical calculations. We have developed an enhanced method for recognizing waveform patterns by integrating this technique with a simple Artificial Neural Network (ANN) architecture, making it particularly suitable for embedded systems with limited memory and computational resources.\\

    In \sects{sec:preproc_algorithm} and \numsect{sec:data_gen}, we propose a pattern recognition problem to distinguish a pulse waveform from other spectral deformed versions of it, for Sinc, Gaussian, and Chirp signals. We use such problems to test, in \sect{sec:param_perf}, the performance of ESSC and SSC feature extraction with different ML models, including Naive Bayes (NB), K-Nearest Neighbors (KNN), Support Vector Machine (SVM), and Multi-Layer Perceptron (MLP). In contrast, we also employed Deep Learning (DL) methods, utilizing raw-waveform data with 256 and 1024 samples, to compare them with the ML techniques.\\
    The results in \fig{fig:bar_plot_Acc} show that at least one of the tested ML methods achieves accuracy comparable to that of the 1D-CNN algorithm (for DL networks) at SNR levels above $20\un{dB}$. In particular, the MLP network using ESSC achieves an accuracy of around 90\% across all the studied signals within the specified SNR range. We observe a significant overall performance reduction when using ML algorithms compared to 1D-CNN at SNR levels below $20\un{dB}$. However, such SNR levels, involving Gaussian White Noise with a standard deviation exceeding 10\% of the signal amplitude, are impractical for most pulsed-signal applications. On the other hand, we observe that the accuracies obtained with the ESSC 30-parameter dataset consistently outperform those with the SSC 4-parameter dataset in nearly all cases, with a significant improvement for the Gaussian signal (due to the very few extrema in the waveform).\\
    Finally, \fig{fig:bar_plot_time} shows that an MLP network is the fastest ML method for classifying signals, approximately 4 times faster than the 1D-CNN with 256 elements. Therefore, using the ESSC pre-processing technique as the feature extraction procedure for an MLP network in pulsed-signal pattern recognition tasks is an excellent alternative to DL methods, with a significantly lower computational effort to build the network (as with all ML methods), as shown in \tabla{tab:train_time_mldl}. Thus, this method constitutes a systematic, fast feature-extraction procedure for these signal types, for which we have experience applying the SSC technique \cite{SegForFarAn13}.\\

    We present the performance measurement of an MLP network with ESSC feature extraction in \sect{sec:res_ex-pr} for the proposed pattern-recognition problems (\sects{sec:mathapp} and \numsect{sec:app_ESSC_ANN}, and \figs{fig:Sinc_fil}, \numfig{fig:Gauss_fil}, and \numfig{fig:Chirp_fil}) under various SNR conditions. From the Confusion Matrix (CM) results in \figs{fig:CM_Sinc}, \numfig{fig:CM_Gauss}, and \numfig{fig:CM_Chirp}, we see that the recognition performance decreases as the noise level increases, as expected. The ESSC methodology shows an excellent performance under an acceptable SNR of $25\un{dB}$. Additionally, such good performance is observed at SNR levels above $20\un{dB}$ for the Sinc-pulse (see \fig{fig:CM_Sinc}) and above $15\un{dB}$ for the Chirp-pulse (where the waveforms exhibit frequency modulation, see \fig{fig:CM_Chirp}). Nevertheless, the Gaussian-pulse recognition problem presents similar waveforms across classes (see \fig{fig:Gauss_fil}); thus, the discrimination process is challenging for the method, and performance decreases rapidly at SNR levels below $15\un{dB}$ (see \fig{fig:CM_Gauss}).\\
    We studied the relevance of each ESSC parameter using the ReliefF MATLAB{\circledR} function. The values shown in \fig{fig:ReliefF} indicate that all 30 parameters are relevant to the discrimination signal process in the proposed examples. Nevertheless, the most relevant parameters differ from one waveform to another, and all the parameters contribute to a general waveform recognition procedure. On the other hand, if the recognition procedure for a waveform fails, we can further extend the method by increasing the number of ESSC parameters and adding higher-order moments, cumulants, derivatives, or integrals to improve (in principle) its accuracy.\\

    Summarizing, we conclude that the proposed ESSC pre-processing technique (with 30 parameters or a further extended version)  comprises a systematic and fast feature extraction method, which can be used as the input procedure for an MLP network in pattern-recognition problems, operating significantly faster and with accuracy comparable to a DL network at practical SNR level conditions, which is particularly useful when the computational capabilities are limited.

\section{Acknowledgement}

This work was supported by CONICET and FONCYT (PICT 2013-2600), Agencia Nacional de Promoci\'{o}n de la Investigaci\'{o}n, el Desarrollo Tecnol\'{o}gico y la Innovaci\'{o}n (MINCYT).
We thank Graciela Corral-Briones for helpful comments and discussions.
\\



\appendix
\section{Description of the most common ML and DL methods}\label{app:descMLDL}

    In this appendix, we briefly detail the main features of various known ML and DL methods.
    We selected the most commonly used ML algorithms \cite{Nitish_Biswas_2022}, including Naive Bayes (NB), K-Nearest Neighbor (KNN), Support Vector Machine (SVM), and Multi-Layer Perceptron (MLP). For the DL approach, we chose a Convolutional Neural Network (CNN). These methods are arranged in order of complexity based on the number of tuning parameters they require. Below are the descriptions of the model characteristics:

    \begin{itemize}
        \item \textit{KNN}
        is one of the simplest classification methods that relies on a distance algorithm, such as the Euclidean metric, to assess the similarity between a query point and the K closest examples in the training dataset. The model's training is significantly influenced by the parameter K; larger values can lead to overfitting, while smaller values may result in underfitting \cite{Vraj_Sheth_2022}. One of its main advantages is its straightforward architecture. However, it is also computationally intensive during the training process and sensitive to noisy data.

        \item \textit{NB}
        is a simple and effective classifier based on Bayes' theorem, assuming that features are independent. Although this assumption often does not hold in real data \cite{I_Rish_2001}, NB can still perform well despite this limitation.

        \item \textit{SVM}
        are kernel-based methods grounded in robust theory and suitable for both binary and multiclass classification. In SVM, input data (features) are transformed from a low-dimensional space to a high-dimensional space to make them linearly separable. This transformation enables the solution of a convex optimization problem that can be addressed analytically \cite{Awad_2015}. During the training phase, the optimization process identifies the hyperplane that best separates the classes within the feature space. This hyperplane is defined by support vectors, which are a subset of the available data that delineate the boundary between the classes. SVMs generally perform well, especially when the number of training samples is small. However, selecting the appropriate kernel functions can lead to high computational costs when working with large datasets, and the performance may decrease when the data is noisy \cite{K_C_Gryllias_2012}. Despite these challenges, SVMs have been widely utilized in conjunction with NB prior to the advent of Convolutional Neural Networks (CNNs).

        \item \textit{MLP}
        is a type of neural network that uses feedforward backpropagation. Its basic structure consists of an input layer that holds the data features, followed by one or more hidden layers, each with its own weights and activation functions. These layers are connected to an output layer, which represents multiclass labels. MLPs are known for their flexibility and scalability, making them effective classifiers for modeling nonlinear relationships. They are also robust to noisy data, especially when appropriate regularization techniques are applied during training \cite{Du_2019}. Accordingly, MLPs are particularly well-suited for large-scale, high-dimensional tasks.

        \item \textit{CNNs}
        typically consist of a high density of layers organized into three main blocks. The first block is the convolutional stage, where the convolution operation extracts and optimizes features during training. In the second block, the pooling stage, a downsampling operation is applied to reduce feature dimensionality. Finally, the last block consists of a fully connected neural network that processes the reduced features; here, the labeling task is accomplished by passing the features through an activation function, similar to traditional Multi-Layer Perceptrons (MLPs) \cite{Chauhan_2018}. Due to the numerous tunable parameters, CNNs can be computationally intensive, leading to long training times. They also tend to require more computational resources than traditional machine learning algorithms, even with the availability of graphical processing units (GPUs) and advanced parallel processing techniques. In this work, a one-dimensional CNN (1D-CNN) is employed, which significantly reduces the computational complexity at the convolutional stage compared to processing two-dimensional image data.
    \end{itemize}


\end{document}